\documentclass[useAMS,usenatbib]{mn2e}
\pdfoutput=1

\usepackage{graphics}
\usepackage{graphicx}
\usepackage{amsxtra}
\usepackage{array,longtable}
\usepackage{journals}
\usepackage{url}

\usepackage{latexsym}
\usepackage{amssymb}

\usepackage{ulem}

\usepackage{color}
\newcommand{\ser}{S\'ersic}

\title
[Mirages in galaxy scaling relations]
{Mirages in galaxy scaling relations
}
\author[A.V.~Mosenkov, N.Ya.~Sotnikova, and V.P.~Reshetnikov]
{A.V.~Mosenkov$^{1,2}$\thanks{E-mail: mosenkovAV@gmail.com},
N.Ya.~Sotnikova$^{1,3}$,
and V.P.~Reshetnikov$^{1,3}$
\\
$^1$St.Petersburg State University, Universitetskij pr. 28, 198504
St.Petersburg, Stary Peterhof, Russia \\
$^2$Central Astronomical Observatory of RAS, Russia\\
$^3$Isaac Newton Institute of Chile, St.Petersburg Branch
}

\begin{document}

\date{Accepted 2014 ???. Received ???; in original form 2013 ???}

\pagerange{\pageref{firstpage}--\pageref{lastpage}} \pubyear{2014}

\maketitle

\label{firstpage}
\begin{abstract}
We analyzed several basic correlations between structural parameters 
of galaxies. The data were taken from various samples in different 
passbands which are available in the literature. We discuss disc 
scaling relations as well as some debatable issues concerning the 
so-called Photometric Plane for bulges and elliptical galaxies in 
different forms and various versions of the famous Kormendy relation.

We show that some of the correlations under discussion are artificial 
(self-correlations), while others truly reveal some new essential 
details of the structural properties of galaxies. Our main results 
are as follows: 

(1) At present, we can not conclude that faint stellar discs are, on 
average, more thin than discs in high surface brightness galaxies. 
The ``central surface brightness -- thickness'' correlation appears 
only as a consequence of the transparent exponential disc model to 
describe real galaxy discs.

(2) The Photometric Plane appears to have no independent physical 
sense. Various forms of this plane are merely sophisticated versions 
of the Kormendy relation or of the self-relation involving the 
central surface brightness of a bulge/elliptical galaxy and the 
\ser\ index $n$.

(3) The Kormendy relation is a physical correlation presumably 
reflecting the difference in the origin of bright and faint 
ellipticals and bulges.

We present arguments that involve creating artificial samples to prove 
our main idea.

\end{abstract}

\begin{keywords}
galaxies: kinematics and dynamics -- galaxies: structure.
\end{keywords}

\section{Introduction}

Global characteristics of galaxies (luminosity, size, rotational 
velocity, velocity dispersion, etc.) are not distributed randomly, 
but form a set of well-defined scaling relations. These relations 
are of great importance since they provide invaluable constraints 
on the formation scenarios and evolutionary processes of galaxies. 
The success of any particular theory will be judged by its ability 
to reproduce the slope, scatter, and zero-point of known scaling 
relations.

One of the most firmly established empirical scaling relations of 
elliptical galaxies is the Fundamental Plane (FP) which represents 
a tight correlation between the surface brightness, the size, and the 
velocity dispersion of a galaxy 
\citep{djorg1987,dressler+1987}. 
Two different projections of the FP are also well-known: 
the Kormendy and the Faber--Jackson relations 
\citep{korm1977a,korm1977b,faber1976}.

Spiral galaxies are more complex since they consist of a disc, a 
bulge, and some other components like, for instance, a bar and a 
ring. The multi-component structure of spiral galaxies results in a 
variety of scaling relations involving parameters associated with a 
disc, a bulge, and a galaxy as a whole. The most famous relation is, 
of course, the Tully--Fisher law \citep{tully1977} which links 
galactic total luminosity and rotation velocity (usually taken as 
the maximum of the rotation curve  well away from the center).

A tight scaling relation may also exist between the photometric and
kinematic characteristics of the discs alone. For instance, 
\citet{kar1989}, \citet{mor+1999}, and others have discussed a 
three-dimensional plane involving the disc scalelength $h$, the 
maximal rotational velocity $v_\mathrm{m}$, and the deprojected 
central surface brightness $S_\mathrm{0,d}$. This plane is analogous 
to the FP of elliptical galaxies.

Edge-on spiral galaxies provide a possibility to analyze the vertical 
surface brightness distribution and, therefore, add a new parameter 
to scaling relations --- the vertical scaleheight, typically, $h_z$ 
for the exponential vertical surface brightness distribution 
\citep{wainscoat+1989} or $z_0$ for the `isothermal' law 
\citep{spitzer1942,vderk1981a,vderk1981b,vderk1982a,vderk1982b}.
A significant correlation between the central surface brightness of 
a stellar disc reduced to the face-on orientation $S_\mathrm{0,d}$ 
and the ratio $h/z_0$ was found: the thinner the galaxy, the fainter 
the central surface brightness \citep{biz2002,biz2004,biz2009}.

Bulges of spiral galaxies also show several empirical relations.
For instance, they follow a relation similar to the FP for elliptical 
galaxies \citep[e.g.][]{falc+2002}.
\citet{kho+2000a} and \citet{kho+2000b} found a tight correlation 
between the \ser\ index $n$, the central surface brightness 
$\mu_\mathrm{0,b}$, and the effective radius of the bulge 
$r_\mathrm{e,b}$. They called this relationship the photometric 
plane (PhP). The PhP projection --- the correlation between the 
central surface brightness of a bulge $\mu_\mathrm{0,b}$ and its 
\ser\ index $n$ --- is known as well 
\citep[e.g.][]{kho+2000b,mol2001,aguerri+2004,ravikumar+2006,barway+2009}. 
Also, there exist mutual correlations between the structural 
parameters of discs and bulges 
(e.g. \citealp{mos+2010}, MSR10 hereafter, and references therein).

In this paper, we critically examine several important scaling 
relations of spiral and elliptical galaxies focusing on spirals, 
their discs and bulges. Our main conclusion is that some of these 
empirical relations (the deprojected central surface brightness -- 
the relative thickness of the disc, the central surface brightness 
of the bulge --- the \ser\ index, the PhP for ellipticals and bulges 
of spirals) are not physical, and they merely reflect the structure 
of fitting formulas. In other words, these scaling relations are 
spurious self-correlations, or mirages of the approximation 
procedures. Such spurious self-correlations arise when two parameters, 
for example, $A$ and $B$ that are used in a linear regression 
analysis, have a common term: $A=f(x)$ and $B=f(x)+g(y)$, where $x$ 
and $y$ are random, uncorrelated variables. In this case, any 
correlation found between $A$ and $B$ has no physical meaning and 
is entirely due to the self-correlation associated with the shared 
variable $x$. Thus, self-correlations link a measured 
parameter $A$ with an expression $B$ including the same parameter. 
Examples are presented to show that under certain conditions perfect 
(but entirely spurious) correlation is obtained between two such 
parameters formed from random numbers. 

On the other hand, we show that the curvature of the Kormendy 
relation is real and can not be explained in terms of other linear 
relations unifying faint and bright galaxies as well as faint and 
bright bulges \citep{grah2003,grah2011,grah2013a}.

This paper is organized as follows. 
In Section~\ref{s_samples}, 
we describe the samples analyzed in this paper. We briefly discuss 
the methods of deriving photometric parameters of bulges and discs.
In Section~\ref{s_corr_discs}, 
we discuss one well-known scaling relation for edge-on discs 
(between the central surface brightness of a stellar disc and the 
relative thickness) and show that it is a self-correlation.
In Section~\ref{s_corr_bulges}, 
we demonstrate that the Photometric Plane for bulges, and ellipticals, 
and its various forms are merely the self-relation involving the 
central surface brightness of a bulge/elliptical and the \ser\ index 
$n$ or sophisticated versions of the Kormendy relation.
In Section~\ref{s_KR}, 
we present arguments in favor of the reality of the Kormendy relation 
which do reveal important features of the galaxy structure.
In Section~\ref{s_conclusions}, we summarize our main conclusions.

\section[]{The samples}
\label{s_samples}

We consider some of the most well-known samples with published 
decomposition results. The samples of edge-on galaxies are provided 
in Table~1. Other selected samples of galaxies are listed in Table~2. 
These samples comprise objects of different morphological types as 
well as are given in different photometric bands. Some samples are 
quite enormous \citep{sim+2011} or huge (e.g., \citealp{allen+2006} 
and \citealp{gad2009}, hereafter G09) whereas others consist of only 
tens objects. We do not consider spheroidal galaxies and ``core'' 
elliptical galaxies since they are out of the scope of this article.

\begin{table*}
 \centering
 \begin{minipage}{150mm}
  \parbox[t]{150mm} {\caption{List of analyzed samples 
  of edge-on galaxies with derived structural parameters of discs.}}
  \begin{tabular}{lrcccccl}
  \hline 
  \hline
 Reference    & Number of galaxies & Band & Morphological types \\
 \hline 
\citet{biz2002} (BM02)     &  134 & $J,H,K_s$ & late types \tabularnewline
\citet{mos+2010} (MSR10)    &  165 & $J$   & all types   \tabularnewline
             &  169 & $H$   & all types   \tabularnewline
             &  175 & $K_\mathrm{s}$ & all types   \tabularnewline
 \hline
\end{tabular}
\end{minipage}
  \label{Table_edge_ons}
\end{table*}

\begin{table*}
 \centering
 \begin{minipage}{150mm}
  \parbox[t]{150mm} {\caption{List of some published samples of 
  galaxies with derived bulge/disc structural parameters of bulges.}}
  \begin{tabular}{lrcccccl}
  \hline 
  \hline
 Reference    & Number of galaxies & Band & Morphological types \\
 \hline 
\citet{caon+1993} &   45 & $B$ & E and S0 \tabularnewline
\citet{maca+2003} &  121 & $B,V,R,H$ & late types  \tabularnewline
\citet{mol2004}   &   26 & $U,B,V,R,I$ & all types \tabularnewline
\citet{allen+2006} (MGC) &  10\,095 & $B$ & all types  \tabularnewline
\citet{sim+2011}  &  1\,123\,718 & $g,r$ & all types \tabularnewline
\citet{gad2009} (G09)   &  946 & $g,r,i$ & all types \tabularnewline
\citet{mcdonald+2011}  &  286 & $g,r,i,z,H$   & all types, Virgo \tabularnewline
\citet{gutierrez+2004} &  187 & $r$ & all types, Coma \tabularnewline
 \hline
\end{tabular}
\end{minipage}
  \label{Table_PhP}
\end{table*}

It should be noted that the structural parameters for these samples 
were derived using various approaches. There are two basic methods: 
the one-dimensional (1D) and the two-dimensional (2D) methods. In the 1D 
method, the azimuthally averaged surface brightness profile of a 
studying galaxy, or major/minor axes profiles, are fitted by one 
or more components. This method has the advantage of being simple 
and fast and works in low signal-to-noise conditions. However, in 
2D fitting, information from the whole image is used to build a more 
robust model for each component. There are several examples in the 
literature showing that the 2D method is much more reliable than the 
1D method \citep[e.g.][]{dejong1996} retrieving more accurate 
structural parameters. 

In this article we do not compare these methods, but rather discuss 
the main results coming from all of them, regardless of the fitting 
procedure.

We should note here that distances to galaxies used by the 
authors were differently estimated for each sample. The Hubble 
constant $H_0$ varies from 70 to 75~km\,s$^{-1}$\,Mpc$^{-1}$ 
what may slightly change the distances. In addition to that, 
for some samples there was no information given on how those 
distances were found, e.g. were the radial velocities of the 
galaxies corrected to the centroid of the Local Group or to the 
galactic center. The vast majority of galaxies from the samples 
are not nearby, and, thus, such corrections do not change 
significantly the distances (in this case the difference may 
variate up to 10\%).

\rm

\section[]{Discs: scaling relations involving scaleheight}
\label{s_corr_discs}

The disc structure out from the galaxy midplane can be investigated 
only for a special galaxy orientation when a disc galaxy is 
seen edge-on to the line of sight. It provides a unique opportunity 
to build a full 3D model of a galaxy and to define the disc thickness. 
Observations of the edge-on galaxies reveal also large-scale 
features that would otherwise remain hidden, like warps, truncations, 
bright halos, and boxy/peanut-shaped bulges. For objects thus 
oriented, one can study the distributions and ages of stellar 
populations. All these issues provide essential insights into 
the formation and evolution of disc galaxies. 

In this section, we focus on the vertical structure of galactic discs 
and on one scaling relation that incorporates the thickness of the 
stellar disc and its deprojected central surface brightness 
(\citealp{biz2002,biz2009}; BM02 and BM09 hereafter). As we use the 
relative thickness of the disc ($z_0/h$), the difference in distances 
to galaxies from different samples does not affect the relations.

\subsection[]{Scaling parameters for edge-on discs}

\begin{figure}
\centering
\includegraphics[width=9.0cm, angle=0, clip=]{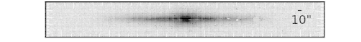}
\includegraphics[width=9.0cm, angle=0, clip=]{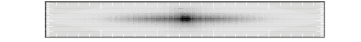}
\includegraphics[width=9.0cm, angle=0, clip=]{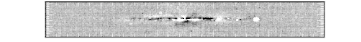}
\caption{Decomposition of the edge-on galaxy NGC~891 
  (2MASS, $K_\mathrm{s}$ band). 
  Images from top to bottom are the galaxy, 
  the model, and the residual image. 
  The dust lane is distinct on the residual image. 
  The derived parameters of the disc are 
  $\mu_{0,d}$=15.8\,mag\,arcsec$^{-2}$, 
  $h$=95.8\,arcsec, $z_\mathrm{0}$=12.9\,arcsec; 
  for the bulge: $\mu_\mathrm{e,b}$=17.7\,mag\,arcsec$^{-2}$, 
  $r_\mathrm{e,b}$=25.65\,arcsec, $n$=2.3, and the apparent bulge axis ratio $q_\mathrm{b}$=0.8. }
\label{NGC891}
\end{figure}

The breakthrough study of edge-on galaxies appeared in the 1980s when 
van der Kruit, Searle 
(e.g., \citeyear{vderk1981a,vderk1982a}), 
and then other authors wrote several 
classical papers on the study of edge-on galaxies. Since that time, 
much progress has been made to investigate these objects 
(e.g., 
\citealp{resh1997,degr1998,kregel+2002,pohlen+2000,pohlen+2004,yoa2006}) 
and to summarize the main conclusions made from previous studies 
(e.g., MSR10; \citealp{vderk2011}). 
Following these studies, we can derive the parameters of
two major stellar components: a bulge and a disc, where the 
disc can be described by the law which comprises the exponential 
radial scale as well as the heightscale (these are 
necessary for building the 3D surface brightness distribution of 
an observed galactic disc): 
\begin{equation}
I(r,z) = 
I(0,0)
\displaystyle \frac{r}{h} \, K_1\left(\frac{r}{h}\right) 
\mathrm{sech}^2(z/z_0) , \,
\label{form_DSB}
\end{equation}
where $I(0,0)$ is the disc central intensity, $h$ is the radial 
scalelength, $z_0$ is the `isothermal' scaleheight of the disc 
\citep{spitzer1942}, and $K_1$ is the modified Bessel function of the 
first order. This formula is valid only in the case of a perfectly 
transparent disc. Unfortunately, the dust within galaxy discs can 
strongly attenuate the light not only from their discs but also from 
the embedded bulges. Dust lanes which are especially prominent in 
early type spiral galaxies (the flocculent dust content often resides 
also in late type spirals) may cover the significant part of the 
galactic disc what can be crucial to correctly determine the disc 
and bulge structural parameters. This effect can be considerable even 
in the NIR bands. For instance, an edge-on galaxy NGC~891 has a dust 
lane that is very visible in the $K_\mathrm{s}$ band 
(see Fig.~\ref{NGC891}). In addition, one of the difficulties we are 
faced with while studying edge-on galaxies, is that we are not able 
to observe a spiral pattern in them. Thus, a guess as to the 
morphological type of a galaxy can be made mainly on the basis of 
its bulge-to-disc luminosity ratio. 

\subsection[]{Central surface brightness~--~thickness relation}

\begin{figure}
\centering
\includegraphics[width=3.0in, angle=0, clip=]{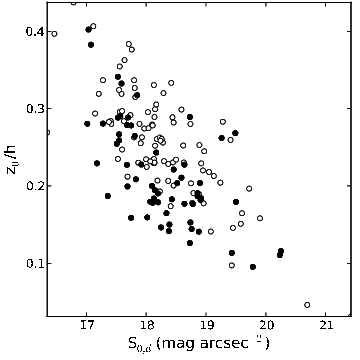}
\caption{Correlation between the relative thickness of a disc 
  and its reduced central surface brightness in the $K_\mathrm{s}$ 
  band. Data were taken from BM02. Black filled circles correspond 
  to the more reliable subsample (designated as ``x'' in the table~1 
  from BM02).}
\label{S0d_biz}
\end{figure}

BM02 analyzed a sample of late-type edge-on galaxies in the 
$J$, $H$ and $K_\mathrm{s}$ bands (see Table~1). 
They have noted a strong 
correlation between the central surface brightness of a stellar disc 
and the $h/z_0$ ratio. This means that the thinner a galaxy is, the 
lower its central surface brightness reduced to the face-on 
orientation $S_\mathrm{0,d}$ (we will designate the apparent central 
surface brightness of edge-on galaxies as $\mu_\mathrm{0,d}$). The 
same correlation was confirmed for the stellar disc structural 
parameters corrected for internal extinction (\citealp{biz2004}; BM09). 
\citet{biz2004} noted that this extinction correction is rather small 
(the median value for their sample is about 0.1 mag/arcsec$^2$). 
Fig.~\ref{S0d_biz} demonstrates the $S_{0,d}$~--~$z_0/h$ correlation. 
BM09 concluded that a very wide scatter of points in this correlation 
is due to the relatively low accuracy in the $\mu_\mathrm{0,d}$, $z_0$ 
and $h$ and is also due to the non-constant value of the mass-to-light 
ratio ($M/L$) for different galaxies.

\subsection[]{How does the relation 
$S_{0,d}$~--~$z_0/h$ reveal itself in other samples}

The largest sample of edge-on galaxies with derived structural 
parameters of discs and bulges is the MSR10 sample (see Table~1). 
It comprises both early and late type objects. The fits-images were 
taken from 2MASS in all three bands ($J$, $H$ and $K_\mathrm{s}$). 
The sample is incomplete according to the $V/V_\mathrm{max}$ test, 
but the subsample of 92 galaxies with angular radius $r>60$~arcsec 
appears to be complete. The program BUDDA 
(Bulge/Disc Decomposition Analysis; \citealp{desouza+2004}) was 
applied for performing bulge/disc decomposition. As we have all needed 
structural parameters, we can construct the same relation as in BM02.

\begin{figure}
\centering
\includegraphics[width=9.0cm, angle=0, clip=]{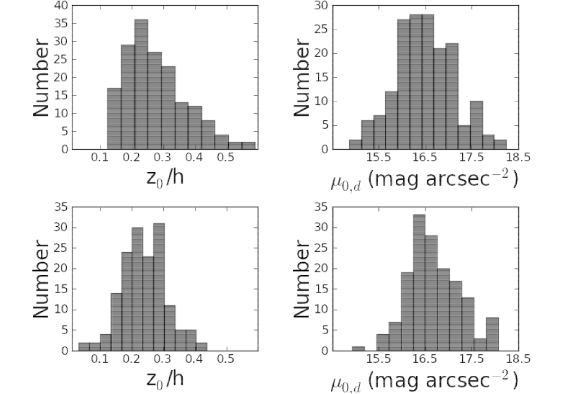}
\caption{Distributions of the relative thickness and the deprojected 
  central surface brighness of the discs in the $K_\mathrm{s}$ band 
  for the sample by MSR10 (top plots) and for the BM02 sample 
  (bottom plots).}
\label{distr}
\end{figure}

Let us compare the sample by MSR10 and the BM02 sample. 
In Fig.~\ref{distr} we plotted the distributions of the parameters 
$z_\mathrm{0}/h$ and $\mu_\mathrm{0,d}$ in the $K_\mathrm{s}$ band. 
The BM02 sample comprises mainly late-type spiral galaxies. 
That is why the distributions over photometric parameters for this 
sample look slighlty different in comparison with our sample, but the 
mean values of both samples are similar. 

The median values and standard deviations for the MSR10 sample are 
the following: 
$$\langle z_\mathrm{0}/h \rangle = 0.25\pm0.11\,,$$
$$\langle \mu_\mathrm{0,d} \rangle = 16.46\pm0.69\,\mathrm{mag\,arcsec}^{-2}\,,$$
For the BM02 sample:
$$\langle z_\mathrm{0}/h \rangle = 0.23\pm0.07\,,$$
$$\langle \mu_\mathrm{0,d} \rangle = 16.62\pm0.56\,\mathrm{mag\,arcsec}^{-2}\,.$$
From these distributions we can see that the scatters of both 
parameters are relatively narrow, and the characteristics of the 
samples are close.

In Fig.~\ref{S0d_all} we show the mutual distribution of 
$\mu_\mathrm{0,d}$ and $z_\mathrm{0}/h$ for our sample 
(row $a$, left plot) and for the BM02 sample (row $b$, left plot). 
Right plots in Fig.~\ref{S0d_all} represent
the $S_\mathrm{0,d}$~--~$z_\mathrm{0}/h$ correlation for our sample 
(row $a$) and for the BM02 sample (row $b$). 
The regression line for the MSR10 sample is
\begin{equation}
S_\mathrm{0,d} = -5.09 \log(z_0/h) + 14.81, \, r=-0.49 \, ,
\label{reg_our}
\end{equation}
and for the BM02 sample is
\begin{equation}
S_\mathrm{0,d} = -5.17 \log(z_0/h) + 14.71, \, r=-0.684 \, .
\label{reg_BM02}
\end{equation}

Correlations for both samples are similar and rather strong, but are 
they real?

It is known from the surface photometry of transparent discs 
that the central surface brightness of the face-on disc (when the 
inclination angle is $i = 0^\circ$) expressed in magnitudes per 
arcsec$^2$, can be reduced from the edge-on (apparent) central surface 
brightness as follows:
\begin{equation}
S_\mathrm{0,d} = \mu_\mathrm{0,d}  - 2.5 \, \log (z_0/h). \,
\label{form_S0d}
\end{equation}

From this expression~(\ref{form_S0d}) we may notice several useful 
facts. First, the scatter of $S_\mathrm{0,d}$ should be larger than 
that of $\mu_\mathrm{0,d}$ because of the presence of the term 
$\log (z_\mathrm{0}/h)$. 
Second, from~(\ref{form_S0d}) we can see that if 
$z_0/h \approx \mathrm{const}$, then there is a simple linear
dependence between $S_\mathrm{0,d}$ and $\mu_\mathrm{0,d}$. 
Third, contrary, if $\mu_\mathrm{0,d} \approx \mathrm{const}$, there 
is a simple logarithmic dependence between $S_\mathrm{0,d}$ and 
$z_\mathrm{0}/h$. Hence, the small scatters around the median values 
$\langle \mu_\mathrm{0,d}\rangle$ and 
$\langle z_\mathrm{0}/h\rangle$ may transform the reduction 
formula~(\ref{form_S0d}) 
into the self-correlation between $S_\mathrm{0,d}$ and $z_0/h$ 
because the expression for $S_\mathrm{0,d}$ contains the term 
of $z_0/h$. To prove this conclusion, we designed some examples.  
They show that under certain conditions, perfect (but 
entirely spurious) correlation is obtained between two parameters 
formed from random distributions.

\subsection[]{Self-relation between central surface brightness and 
thickness: artificial samples}

We generated several samples of artificial galaxies with normal 
distributions of observed parameters $\mu_\mathrm{0,d}$ and 
$z_\mathrm{0}/h$. Although the distributions of $\mu_\mathrm{0,d}$ 
and $z_\mathrm{0}/h$ are not normally distributed in reality 
(see Fig.~\ref{distr}), we use this simplification merely to show that 
the resultant correlation will be the same as that for the real data. 

\begin{figure*}
\centering
\includegraphics[height=15.0cm]{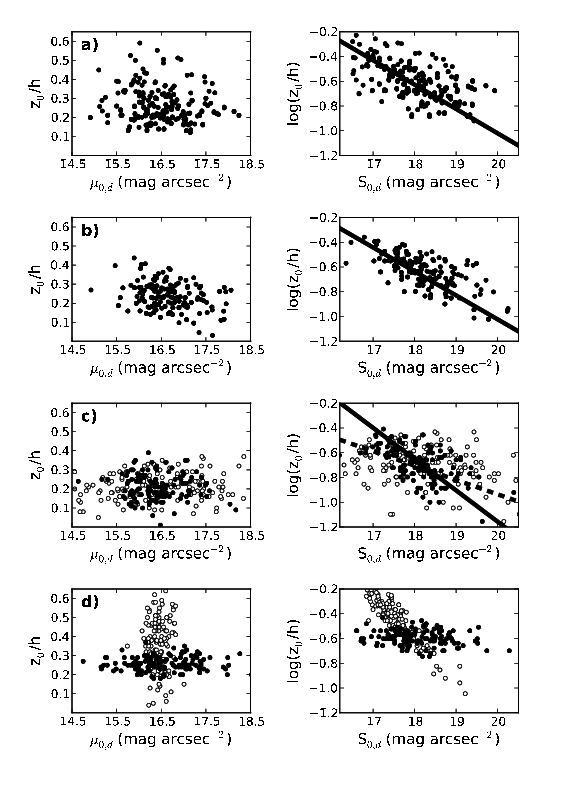}
\caption{Correlations between the relative thickness and the 
  central surface brightness ($K_\mathrm{s}$ band): 
  apparent (left column) and reduced (right column); 
  a) the sample by MSR10; 
  b) the BM02 sample; 
  c) artificial samples~\#1 and~\#2 (filled circles and 
  open circles respectively);
  d) artificial samples~\#3 (filled circles)
  and~\#4 (open circles), see the text. Solid lines correspond 
  to the regression lines for the samples mentioned in the text. 
  Dashed line is a regression line for the sample~\#2.}
  \label{S0d_all}
\end{figure*}

The sample~\#1 (filled circles in Fig.~\ref{S0d_all}c, left plot) 
is built to imitate the real distribution similar to our and the BM02 
samples with the following mean value of $\mu_\mathrm{0,d}$ and its 
standard 
$\sigma$:
\[
\bar{\mu}_\mathrm{0,d}=16.5,\,\sigma=0.6\, \mathrm{mag\,arcsec}^{-2}\,.
\]

The sample~\#2 (open circles in Fig.~\ref{S0d_all}c, left plot) 
has a wider distribution over $\mu_\mathrm{0,d}$:
\[
\bar{\mu}_\mathrm{0,d}=16.5,\,\sigma=1.1\, \mathrm{mag\,arcsec}^{-2}\,.
\]
In both cases the distribution of $z_\mathrm{0}/h$ was the same:
\[
\overline{z_\mathrm{0}/h}=0.25,\,\sigma=0.05\,.
\]

We converted $\mu_\mathrm{0,d}$ into $S_\mathrm{0,d}$ according 
to (\ref{form_S0d}) and plotted the relation 
$S_\mathrm{0,d}$~--~$\log (z_0/h)$ (see the right column in 
Fig.~\ref{S0d_all}). It appears to be linear with a scattering 
that is due to the scatter of $\mu_\mathrm{0,d}$ and $z_\mathrm{0}/h$. 
The regression line for the sample~\#1 is
\begin{equation}
S_\mathrm{0,d} = -3.96 \log(z_0/h) + 15.42, \, r=-0.545 \, , 
\label{reg_s1}
\end{equation}
and the regression line for the sample~\#2 
(with a wide distribution over $\mu_\mathrm{0,d}$) is
\begin{equation}
S_\mathrm{0,d} = -1.41 \log(z_0/h) + 17.17, \, r=-0.21 \, . 
\label{reg_s2}
\end{equation}
The regression coefficient is much smaller for this broader 
distribution of $\mu_\mathrm{0,d}$ 
(the sample~\#2, Fig.~\ref{S0d_all}c, right plot, open circles, 
dashed line). 
Thus, the correlation $S_\mathrm{0,d}$~--$z_0/h$ for this sample 
is statistically insignificant. 
But for the artificial sample~\#1 containing random (uncorrelated) 
distributions of $\mu_\mathrm{0,d}$ and $z_0/h$ the regression 
coefficient and the slope of the correlation 
$S_\mathrm{0,d}$~--$z_0/h$ 
(Fig.~\ref{S0d_all}c, right plot, filled circles, solid line)
are almost the same as that for the BM02 and MSR10 samples.
{\it Thus, we can see that even if we have no 
correlation between the visible surface brightness of the edge-on disc 
and its relative thickness, there would be, nevertheless, the 
correlation between the reduced central surface brightness and the 
relative thickness of the disc.} 

This correlation, however, can be substantially smoothed and even 
destroyed if the scatter of $\mu_\mathrm{0,d}$ is rather large. 
To demonstrate this fact, we constructed two additional artificial 
samples. 
The sample~\#3 (filled circles in Fig.~\ref{S0d_all}d, left plot) 
has a very small scatter of the ratio $z_\mathrm{0}/h$ ($\sigma=0.05$) 
and a large scatter of $\mu_\mathrm{0,d}$ (as for the sample~\#1) 
with the same mean values. 
The sample~\#4 (open circles in Fig.~\ref{S0d_all}d, left plot), 
contrary, has a very small scatter of the value 
$\mu_\mathrm{0,d}$ ($\sigma=0.2$) and a wide distribution over the 
ratio $z_\mathrm{0}/h$ ($\sigma=0.1$). 
As a consequence, the sample~\#3 do not show any correlation between 
$S_\mathrm{0,d}$ and $z_0/h$ 
(filled circles in Fig.~\ref{S0d_all}d, right plot). 
In other words, if there is a narrow scatter of $z_\mathrm{0}/h$ 
with the same distribution of $\mu_\mathrm{0,d}$ as for the 
sample~\#1, then the $S_\mathrm{0,d}$~--~$z_\mathrm{0}/h$ correlation 
does not appear. 

On the contrary, if we have a large scatter of $z_\mathrm{0}/h$ 
with a very narrow distribution of $\mu_\mathrm{0,d}$, then the 
expected correlation will be very strong 
(open circles in Fig.~\ref{S0d_all}d, right plot).

Trying to explain the $S_\mathrm{0,d}$~--~$z_0/h$ correlation, \citet{biz2004} noted that the values of $S_\mathrm{0,d}$ and $z_0/h$ 
had not been obtained independently from each other as can be 
concluded from the Eq.~(\ref{form_S0d}). They considered several 
effects that can affect the correlation $S_\mathrm{0,d}$~--~$z_0/h$. 
In particular, they argue 
that a non 90 degree inclination of the disc simply shifts the data 
points in Fig.~\ref{S0d_all} towards the upper right corner because 
of the overestimation of $z_0$. Hence, systematic errors due 
to inclination may only scatter the dependence shown in 
Fig.~\ref{S0d_all} and do not affect a correlation if it exists. 
This explanation can not be adopted because there is certainly a 
scatter around the relation~(\ref{form_S0d}) with the median value of 
$\langle\mu_\mathrm{0,d}\rangle$. The slope of the regression 
lines~(\ref{reg_our}) and~(\ref{reg_BM02}) is twice as large as the 
slope of the relation~(\ref{form_S0d}) with the median value 
$\langle \mu_\mathrm{0,d}\rangle$, but the slope of the relation for 
artificial sample~\#1 is also larger 
(see the expression~(\ref{reg_s1})). In other words, we can not 
assert the existence of the 
$S_\mathrm{0,d}$~--~$z_\mathrm{0}/h$ correlation beyond the 
self-correlation due to the reduction procedure~(\ref{form_S0d}).

\subsection[]{Are there physical bases of the correlation 
$S_\mathrm{0,d}$~--~$z_\mathrm{0}/h$?}

Let us turn to the possible explanation of the correlation 
$S_\mathrm{0,d}$~--~$z_\mathrm{0}/h$, if it exists. 
Following BM09 (see also \citealp{zas+2002,kregel+2005,sot2006}),
we will consider the exponential disc which is in equilibrium in 
the vertical direction. 
For such a disc we can find the vertical scaleheight $z_0$
via the vertical equilibrium condition for an isothermal slab 
\citep{spitzer1942}: 
$\sigma_z^2 = \pi G \Sigma z_0$, where $\sigma_z$ is the vertical 
velocity dispersion, and $\Sigma$ is the surface density of a slab. 
To express the central surface density through the central surface 
brightness, we can write: $\Sigma_0 \propto 10^{-0.4S_\mathrm{0,d}}$. 
The mass of the disc can be estimated as 
$M_\mathrm{d}=2\pi \Sigma_\mathrm{0}h^2$. 
At $R \simeq 2h$, the rotation curves of luminous galaxies generally 
reach a plateau. In the plateau region, the linear circular velocity
$v_\mathrm{c}$ is roughly constant. We can then use $v_\mathrm{c}$ to
estimate the total mass of a galaxy (including the mass of its 
spherical component: bulge$+$dark halo) within the sphere of the 
radius $R=4h$: $M_\mathrm{tot}(4h)=4 v_\mathrm{c}^2h/G$. 
Thus, we have:
\begin{equation}
\frac{M_\mathrm{tot}(4h)}{M_\mathrm{d}} \propto 
\frac{v_\mathrm{c}^2}{h} 10^{0.4S_\mathrm{0,d}} \, .
\label{MtotMdisc}
\end{equation}

We can link the relative mass of a disc with its relative thickness 
via stability conditions.
If the stellar disc is marginally stable in its plane, 
then the radial velocity dispersion can be written as 
$\sigma _R(R) = Q \, \displaystyle \frac{3.36G\,\Sigma(R)}{\kappa(R)}$, 
where $Q$ is the Toomre parameter \citep{toomre1964}, 
$\kappa$ is the epicyclic frequency, $R$ is the 
radius in the cylindrical reference frame associated with the disc. 
For marginally stable discs the radial profile of $Q$ usually has 
a wide minimum with the value $Q \approx 1.5$ in the region of 
$(1-2)\cdot h$. This value is justified by the results of numerical 
experiments by \citet{khop+2003}. Thus, we can consider $Q$ to be 
almost constant with the radius outside the disc center.
The epicyclic frequency at the region where 
$v_\mathrm{c} \approx \mathrm{const}$, is 
$\kappa \approx \sqrt{2} v_\mathrm{c}/R$. 
In summary, we obtain
for\footnote{The reference distance of $R \approx 2h$ is chosen 
because $Q$ and $v_\mathrm{c}$ are almost constant there.} 
$R \approx 2h$ (see \citealp{sot2006} for details):
\begin{equation}
\frac{h}{z_0}
 \propto 
 \frac{1}{\left(\sigma_z/\sigma _R\right)^2}
 \frac{v_\mathrm{c}^2/h}{\Sigma}
 \propto 
 \frac{1}{\left(\sigma_z/\sigma _R\right)^2}
 \frac{M_{\mathrm{tot}}(4h)}{M_\mathrm{d}} 
\, .
\label{z0h_MtMd}
\end{equation}
If the ratio of vertical to radial velocity dispersions 
$\sigma _z/\sigma _R$ is almost constant throughout the disc, we have 
a correlation between $z_\mathrm{0} / h$ and 
$M_\mathrm{d}/M_\mathrm{tot}$. The existence of such a correlation 
was for the first time mentioned by \citet{zas+1991}, and it was 
further explored in many works 
(e.g. \citealp{zas+2002,kregel+2005,sot2006}; MSR10) 
and references in BM09). 
The ratio $\sigma _z/\sigma _R$ could be fixed at the level given by
the local linear criterion for the marginal bending stability, i.e.
at approximately 0.3 
(\citealp{toomre1966,kulsrud+1971,pol1977,ara1985}). Recent numerical 
experiments by \citet{rod2013} support this minimal value throughout 
the disc. For real galaxies, some mechanisms heating the disc in 
the vertical direction and causing an increase in the ratio 
$\sigma _z/\sigma _R$ may operate. At present, the ratio 
$\sigma _z/\sigma _R$ is measured directly only in a few 
galaxies \citep{gerssen+1997,gerssen+2000,shapiro+2003,gerssen2012}. 
It ranges from 0.3 to 0.8, but for our purposes we can fix this value at 
any level. 

\begin{figure}
\centering
\includegraphics[width=8.0cm]{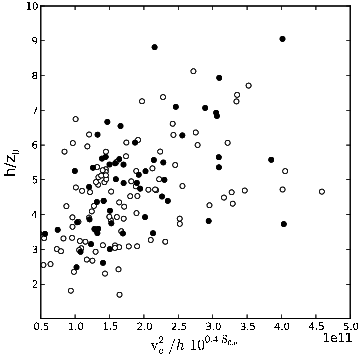}
\caption{Correlation between $z0/h$ and 
$v_\mathrm{c}^2/h\, 10^{0.4S_\mathrm{0,d}}$ 
  (which is proportional to $M_\mathrm{d}/M_\mathrm{tot}$) 
  in the $K_\mathrm{s}$ band. 
    Open circles represent the complete subsample of the sample by 
  MSR10, filled circles represent the BM02 subsample. 
  The rotational velocity values $v_\mathrm{c}$ were taken from 
  the LEDA database (as uncorrected for inclination \texttt{vmax} 
  output parameters supposing that the inclination 
  $i\approx90^\circ$ for all galaxies considered).}
\label{z0_v2S0d}
\end{figure}

Now, combining~(\ref{MtotMdisc}) and~(\ref{z0h_MtMd}) we can expect:
\begin{equation}
\frac{h}{z_0} \propto 
\frac{v_\mathrm{c}^2}{h} 10^{0.4S_\mathrm{0,d}} \, .
\label{z0h_S0d}
\end{equation}
BM09 came to a similar conclusion. They used the correlation 
$h \sim v^{1.5}$ which they had observed, and found 
$z_\mathrm{0} / h \sim \Sigma_0 / h^{1/3}$, 
where $\Sigma_0$ is the central surface density of a disc. 
They considered such a result to be the theoretical basis for their 
correlation between $h/z_0$ and $S_\mathrm{0,d}$ (BM02). 
We need to emphasize, however, that in the relation~(\ref{z0h_S0d}) we 
have not only the term $S_\mathrm{0,d}$ but also $v_\mathrm{c}^2/h$ 
(in the BM09' version it is $h^{1/3}$). From this correlation it is 
not evident that there is a correlation between $S_\mathrm{0,d}$ and 
$(z_0/h)$ only! We may only conclude that there may be a 
correlation~(\ref{z0h_S0d}) which actually takes place as we can see 
in Fig.~\ref{z0_v2S0d}. Correlation~(\ref{z0h_S0d}), however, 
comprises the $S_\mathrm{0,d}$ term which was received via 
relation~(\ref{form_S0d}). Correlation~(\ref{z0h_S0d}), therefore, 
exists in the same  sense as the correlation 
$S_\mathrm{0,d}$---$z_0/h$ exists. We can not prove the reality of 
this correlation for galaxies at moderate inclination for which 
$S_\mathrm{0,d}$ can be derived directly without reduction formula. 
Unfortunately, for such galaxies the ratio $z_0/h$ is undefined.

Moreover, the correlation between $z_0/h$ and 
$M_\mathrm{d}/M_\mathrm{tot}$ 
that was used to come to~(\ref{z0h_S0d}), is rather ambiguous, mainly 
due to the term $(\sigma_z/\sigma _R)^2$ in the 
expression~(\ref{z0h_MtMd}) (\citealp{sot2006}; MSR10). 
It exists only in the sense that discs embedded into very massive 
halos are always very thin.

\subsection[]{Conclusion}

The correlation between $h/z_0$ 
and $S_\mathrm{0,d}$, if it exists, is rather weak and can not be 
derived from observational data because the main effect seen in 
Fig.~\ref{S0d_all} is predominantly due to data reduction, many 
assumptions, and specific mathematical laws used to describe disc 
surface brightness distribution. All together, it gives a predictable 
result, i.e. a self-correlation.

\section[]{Bulges and ellipticals: Photometric Plane}
\label{s_corr_bulges}

\subsection[]{Background}

The overall shape of elliptical, dwarf elliptical and bulge profiles 
can be quantified and parametrized by means of $r^{1/n}$ law 
\citep{ser1968} for the radial surface brightness $I(r)$ which is a 
simple generalization of $r^{1/4}$ \citep{dev1948,dev1953,dev1959} 
and exponential laws by \citet{freeman1970} 
(see, for example, \citealp{davies+1988,young1994,grah2001} 
and references therein). 

The $r^{1/n}$ profile is given by the formula:
\begin{equation}
 I(r) = I_0 \, 
 e^{-\nu_n(r/r_\mathrm{e})^{1/n}}\, ,
\label{SBr}
\end{equation}
where $r_\mathrm{e}$ is the effective radius, i.e. the radius of the 
isophote that contains 50\% of the total luminosity of a galaxy or a 
bulge, $I_0$ is the central surface brightness, $n$ is the \ser\ index 
defining the shape of the profile, and the parameter $\nu_n$ ensures 
that $r_\mathrm{e}$ is the half-light radius. In magnitudes per 
arcsec$^2$ the expression (\ref{SBr}) appears as follows
\begin{equation}
 \mu(r) = \mu_0 + 
 \frac{2.5\, \nu_n}{\ln{10}} \, 
 \left(\frac{r}{r_\mathrm{e}}\right)^{1/n}\, ,
\label{form_mu0}
\end{equation}
where $\mu_0$ is the central surface brightness expressed in mag per 
arcsec$^2$. The coefficient $\nu_n$ depends on $n$ and is an almost 
linear function of the \ser\ index $n$. As usual, one implies a numerical 
approximation of $n$ in any appropriate form. One of these 
approximatione which is valid in the range 
$0.5 \leq n \leq 16.5$, is \citep{caon+1993}
\begin{equation}
\frac{\nu_n}{\ln{10}} \simeq 0.868 \, n-0.142 \, .
\label{bn}
\end{equation}

The profile of an elliptical galaxy (and a bulge of a spiral) 
that is fitted with the \ser\ model, can be also expressed as
\begin{equation}
\displaystyle
\mu(r) = \mu_\mathrm{e} + 
1.0857 \, \nu_n \left[ \left(r/r_\mathrm{e}\right)^{1/n}-1\right] \, .
\label{form_mue}
\end{equation}
where $\mu_\mathrm{e}$ is the effective surface brightness, i.e. 
the surface brightness at $r_\mathrm{e}$.


For the fitting purpose, we can use the formula~(\ref{form_mu0}) and 
consider $\mu_0$ and $n$ as free (fit) parameters fixing the range 
of possible values of $\mu_\mathrm{e}$ \citep[e.g.][]{caon+1993}. 
In this case the uncertainty associated with the determination of 
$\mu_\mathrm{e}$ arises because 
$\mu_\mathrm{e}^{*} = \mu_0 + 1.0857 \, \nu_n$ can differ from its 
measured value $\mu_\mathrm{e}$. The value of $\mu_\mathrm{e}^{*}$ 
can be further compared with the measured counterpart 
$\mu_\mathrm{e}$ to test the goodness of a fit. 
On the contrary, if a fit for a sample is ambiguous and comprises 
systematic errors, such errors may affect scaling relations.

In the bulge-disc decomposition, we have the following 
as free (fit) scaling parameters for a bulge: 
(1) the central bulge intensity $I_\mathrm{0,b}$ in counts what can 
be later converted to $\mu_\mathrm{0,b}$ in mag arcsec$^{-2}$; 
(2) the half-light radius of the bulge $r_\mathrm{e}$ in pixels; 
(3) the bulge shape parameter $n$ \citep[e.g.][]{kho+2000b,kho+2004}. 
In this case, $\mu_\mathrm{e,b}$ can be calculated from the expression
\begin{equation}
\mu_\mathrm{e,b} = \mu_\mathrm{0,b} + 1.0857 \, \nu_\mathrm{n} \, .
\label{form_mueb_mu0b}
\end{equation}

It has become customary to choose $\mu_\mathrm{e,b}$ as a fit 
scaling parameter instead of $\mu_\mathrm{0,b}$ 
(\citealp{mol2001,maca+2003,bal+2003,men+2008}; G09; MSR10). 
In this case, $\mu_\mathrm{0,b}$ is not an independent parameter 
but is calculated from the formula~(\ref{form_mueb_mu0b}) that 
involves~$n$.

\subsection[]{Photometric Plane as a bivariate relation}
\label{ss_PhP}

The derived scaling parameters of galaxies may correlate. 
Correlations comprising the scaling parameters of \ser\ models, 
are widely discussed in the literature as well as the physical 
reasons of such correlations. 
\citet{grah2003} discussed several linear scaling relations for 
elliptical galaxies (mainly for dEs and intermediate to bright E 
galaxies). There are also bivariate correlations. One of them 
was introduced by \citet{kho+2000a} and \citet{kho+2000b} and was 
called Photometric Plane (PhP). 
Many authors have confirmed it for their samples of elliptical galaxies and 
bulges of spiral galaxies of all types in various bands 
\citep{mol2001,ravikumar+2006,men+2008,lauri+2010}, 
in different environments \citep{kho+2004}, and 
for faint and bright objects \citep{barway+2009}.

\citet{kho+2000b} presented the PhP as a bivariate relation that links 
only photometric parameters obtained by fitting a \ser\ model to a 
galaxy image (or to a bulge), i.e. the \ser\ index $n$, the central 
surface brightness\footnote{Hereafter we denote the surface brightness 
for ellipticals and bulges as $\mu_\mathrm{0,b}$ or $\mu_\mathrm{e,b}$ 
to distinguish it from the surface brightness of discs.} 
$\mu_\mathrm{0,b}$, and the effective radius of a 
galaxy, or of a bulge $r_\mathrm{e,b}$.

For any sample we can perform the least-squared fit of the expression 
\begin{equation}
\log(n) = a \log (r_\mathrm{e,b}) + b \, \mu_\mathrm{0,b} + c
\label{form_PhP1}
\end{equation}
and find $a$, $b$ and $c$\footnote{Here and below we use \texttt{lm} 
function in R language to calculate coefficients of the model.}. 
In the literature there are different versions of the 
Photometric Plane, and we refer to the plane in the 
form~(\ref{form_PhP1}) as the PhP1.

\citet{kho+2000b} concluded that there exist two univariate 
correlations between the effective radius and the \ser\ index $n$, 
and between $n$ and the central surface brightness. 
These univariate correlations have a scatter that may be caused by a 
third parameter. The methods of multivariate statistics applied to 
the three parameters $n$, $\mu_\mathrm{0,b}$, and $r_\mathrm{e,b}$ 
may reduce the scatter and give the best-fit plane like that 
expressed by Eq.~(\ref{form_PhP1}).

The Photometric Plane is thought to be a counterpart of a plane of 
a constant specific entropy of galaxies introduced by 
\citet{limaneto+1999}. \cite{limaneto+1999} proposed two laws that 
elliptical galaxies and bulges of spirals must obey if they form and
reach a quasi-equilibrium stage solely under the influence of 
gravitational processes. The first law is the virial theorem, and 
the second one is that a system in equilibrium is in a maximum 
entropy configuration. 

\cite{marquez+2000,marquez+2001} argued that 
after violent relaxation spherical systems may be considered to be 
in a quasi-equilibrium stage. In this stage, the two above mentioned 
laws are valid, and they lead to quasi-constant specific entropy. 
\cite{ravikumar+2006} expressed the specific entropy $S$ in a 
convenient form via three photometric parameters $\mu_\mathrm{0,b}$, 
$r_\mathrm{e,b}$ in kpc, and $n$. If $S = \mathrm{const}$, there exists 
the relation that connects only $\mu_\mathrm{0,b}$, $r_\mathrm{e,b}$ 
in kpc, and $n$. This relation gives the surface (plane) of a constant 
specific entropy. The value of specific entropy may be adjusted so 
to match the specific entropy plane with the Photometric Plane 
(see, for example, \citealp{kho+2004,ravikumar+2006}). 
Such a coincidence between two planes is thought to give a physical 
interpretation of the PhP1. The PhP1 may be understood as a consequence 
of the two laws mentioned above. A physical interpretation of the PhP1 
given by \cite{marquez+2001}, clarifies the processes that drove the 
formation and evolution of galaxies and proves that the PhP1 is not 
simply an artifact of the definitions of the photometric parameters.


\subsection[]{Photometric Plane 1: is it flat?}

In previous papers (MSR10; \citealp{sot+2012}), we revealed that the 
PhP1 in $J$, $H$ and $K_\mathrm{s}$ bands appeared not to be 
flat. It has a prominent curvature towards small values of $n$ 
(with $\log (n) < 0.2$). Such a curvature is not seen in early papers 
that used small samples with rather large values of $n$ 
\citep{kho+2000a,kho+2000b,mol2001}, but it was noticed later 
\citep{kho+2004,ravikumar+2006,barway+2009} and discussed in terms of 
a curved specific entropy surface. We consider the reason for this 
curvature to be quite a different one, and it helps to understand the 
origin of the relation~(\ref{form_PhP1}).

\begin{figure}
\centering
\includegraphics[width=3.5in, angle=0, clip=]{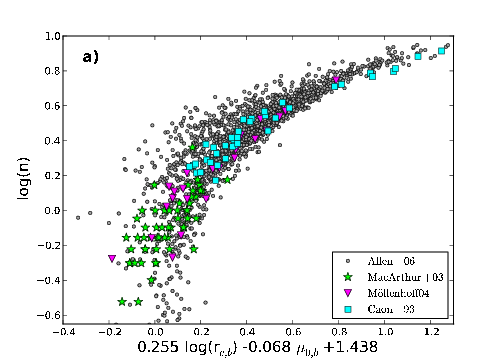}
\includegraphics[width=3.5in, angle=0, clip=]{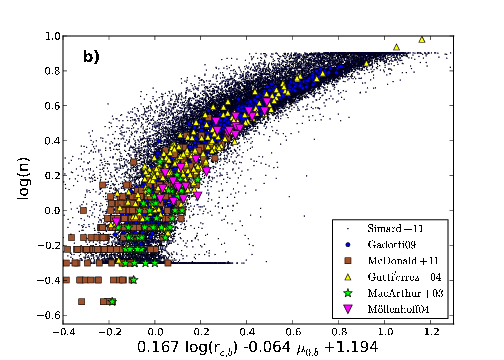}
\caption{The Photometric Plane (PhP1) constructed for 
  a) the MGC sample \citep{allen+2006} in the $B$ band; 
  b) the subsample from \citet{sim+2011} in the $r$ band.
  Some other samples were superimposed on these planes (see the text 
  and the legend).}
\label{PhP1}
\end{figure}

To clarify the question, we found the coefficients of the
expression~(\ref{form_PhP1}) and constructed the PhP1 
(Fig.~\ref{PhP1}a) 
in the $B$ band for more than 10\,000 galaxies of all types from the 
Millennium Galaxy Catalogue --- MGC 
\citep{allen+2006}\footnote{We use the catalogue of structural 
parameters \texttt{mgc\_gim2d.par} from 
\textit{http://www.eso.org/$\sim$jliske/mgc/} recommended by authors. 
We select galaxies with total model magnitude 
$m(\mathrm{B})<19$~mag and $r_\mathrm{e,b}>0.1$~kpc.}.
We superimposed on this plane 45 intermediate to bright E galaxies from \citet{caon+1993}. 
All these samples contain values of $\mu_0$, 
either fitted or recalculated from the model. 
We also added two samples containing structural parameters for bulges 
in the $B$ band: 
121 face-on galaxies of late types from \citet{maca+2003} and 
26 non-barred bright galaxies of all types from \citet{mol2004}.
We did not consider dwarf galaxies because their structure 
can differ substantially from the structure of bright ellipticals and 
bulges. At the same time, we consider bulges and elliptical galaxies 
all together. We will superimpose them often on the same plots 
keeping in mind that these are physically different objects. 
We are inetersted here only in the studying the shape of constructed 
correlations and dependencies which, as we could make sure, are similar for 
both elliptical galaxies and bulges from different samples (despite 
of their possible shift relative to each other on the plotted 
graphs).

We have mentioned that authors used different methods to calculate the 
distances to galaxies. The difference in distances may variate up to 10\% 
and results in a slight difference of physical size of a galaxy. But when 
we compare different samples, we will consider the logarithm of the 
physical size of a component(for example, the effective radius of the bulge), 
so the scatter of its values for all the samples will be less in this scale 
and would not affect the shape of the relation. 

We constructed the same relation (Fig.~\ref{PhP1}b) in the $r$ band 
for the sample from \citet{sim+2011}. The data for 
this sample comes from the Legacy area of the Sloan Digital Sky Survey 
Data Release Seven. This sample contains more than 1 million galaxies, 
sometimes very distant to be analyzed. Therefore we selected objects 
only with $0.02\leq z \leq 0.07$ (more than 200\,000 galaxies). 
To avoid the presence of too many data points on our plots, we 
randomly 
selected 30\,000 galaxies from the subsample. We also used a large 
sample of spiral and elliptical galaxies (946 objects) built by G09 
where galaxies were originally selected with the same restriction on 
$z$. We added the sample of galaxies from the Virgo cluster 
(286 galaxies, \citealp{mcdonald+2011}) and 
187 galaxies in a region of the Coma cluster (mainly Coma cluster 
members, \citealp{gutierrez+2004}). We added to these data 43 galaxies 
with known structural parameters from \citet{maca+2003} and galaxies 
from \citet{mol2004}. 
All samples used in our analysis are listed in Table~2.

It turns out that for all samples there is a fairly 
tight correlation for bulges with $n \ga 2$ (classical bulges) and 
ellipticals, and a big scatter of points for bulges with $n \la 2$ 
(pseudobulges). The curvature of the PhP1 is also quite visible. 
The reason for the curvature may lie in the different nature 
of objects with $n \la 2$ and $n \ga 2$, or in something else. 

\subsection[]{Univariate relations}

As was noted above, the bivariate correlation helps to diminish 
the scatter in univariate correlations. In our case, they are 
correlations between the \ser\ index $n$ and the central surface 
brightness, and between the effective radius and $n$. It is important 
to stress that a narrow plane connecting photometric parameters 
reveals itself only if the expression~(\ref{form_PhP1}) comprises 
the central surface brightness $\mu_\mathrm{0,b}$. A corresponding 
plane does not appear if one uses $\mu_\mathrm{e,b}$ instead of 
$\mu_\mathrm{0,b}$ (see, for example, \citealp{mol2001}). 
Let us now consider two mentioned univariate relations.

\subsubsection[]{Central surface brightness vs \ser\ index for bulges 
and ellipticals}

\citet{grah2003} presented a tight linear relation 
between $\mu_0$ and $\log (n)$ (their figure 9f). The data have been 
compiled from several samples of elliptical galaxies 
\citep{caon+1993,bingg1998,stiav+2001,grah2003}
with derived structural parameters in the $B$ band 
(the \ser\ model was used). Such a 
correlation was noted as a very strong while analyzing data for 
elliptical dwarfs in the Coma cluster \citep{bingg1998,kourkchi+2012}.

Modeling the bulges of spiral galaxies, other authors have
found a similar trend 
\citep[e.g.][]{kho+2000b,mol2001,aguerri+2004,ravikumar+2006,barway+2009}. 

To explain this trend for dEs, \citet{grah2011,grah2013a} discussed two 
key empirical linear relations from 
which the linear relation between $\mu_0$ and $\log (n)$ can be derived. 
They are the luminosity-concentration ($L-n$) relation 
and the luminosity-central density ($L-\mu_0$) relation which unify 
faint and bright elliptical galaxies along one linear sequence. 
This issue will be discussed in Section~\ref{s_KR}.

It should be noted that the points in figure~9f from \citet{grah2003}
do go along a straight line, and the scatter looks natural because of 
inhomogeneity in the compiled sample and uncertainties while fitting 
photometric profiles. The deviation from the straight line 
is seen only at small values of $n$, but the sample in this range is 
poor (see also \citealp{grah2011}, figure 2b).

\begin{figure}
\centering
\includegraphics[width=3.5in, angle=0, clip=]{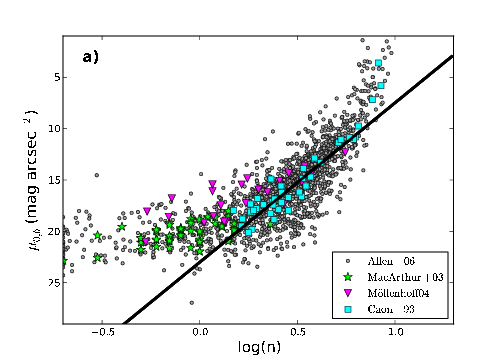}
\includegraphics[width=3.5in, angle=0, clip=]{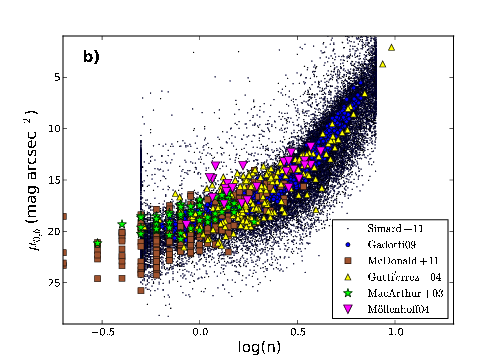}
\caption{The central surface brightness 
  $\mu_\mathrm{0,b}$ of the underlying host galaxy bulge or of the 
  elliptical shown against the \ser\ index 
  $n$ on a logarithmic axis; 
  a) the data are in the $B$ band where the solid line corresponds to the relation $\mu_0 = 23 - 15.5 \, \log(n)$ from \citet{grah2011}; 
  b) the data were taken in the $r$ band (the data of G09 are reproduced separetly in Fig.~\ref{m0b_lgn_grah} and in Fig.~\ref{m0b_n_grah}, where they are plotted with $n$ instead of $\log(n)$).  
  The samples used are listed in Table~2.}
\label{mu0blgn}
\end{figure}

We reproduced the relation $\mu_0$~--~$\log (n)$ in the $B$ band 
(Fig.~\ref{mu0blgn}a) and in the $r$ band (Fig.~\ref{mu0blgn}b) 
for all samples as in Fig.~\ref{PhP1}.
The line $\mu_0 = 23 - 15.5 \, \log(n)$ in Fig.~\ref{mu0blgn}a
was drawn as in \citet{grah2011} where it has been estimated 
by eye. We also reproduced the relation 
$\mu_0$~--~$\log (n)$ in the $r$ band separately for the sample by G09 in the Fig.~\ref{m0b_lgn_grah} while discussing the Kormendy relation. For both bands the relation $\mu_0$~--~$\log (n)$ clearly curves towards the range of small values of $n$ and does not follow a straight line.

\subsubsection[]{Discussion and explanation}

Several questions arise. Why is this relation curved? Is it real? 
Does it reflect some common physical processes which make spherical 
galaxies and bulges acquire their structure? Or, on the contrary, 
can this relation be explained simply by the procedure of image 
decomposition and surface brightness profile fitting?

\begin{figure}
\includegraphics[width=3.5in, angle=0, clip=]{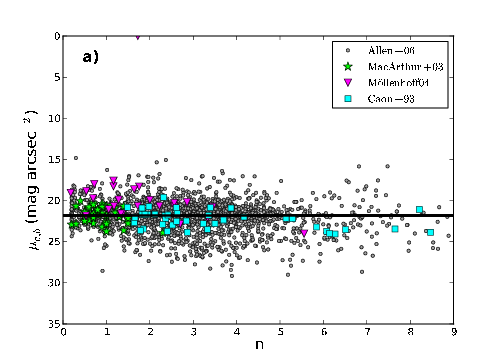}
\includegraphics[width=3.5in, angle=0, clip=]{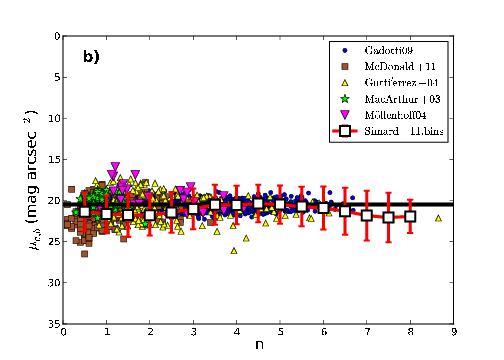}
\caption{The effective surface brightness 
  $\mu_\mathrm{e,b}$ of the underlying host galaxy bulge or of the 
  elliptical shown against the \ser\ indices $n$; 
  a) the data are in the $B$ band, 
  $\langle\mu_\mathrm{e,b}\rangle=21.81$~mag\,arcsec$^{-2}$ 
  corresponds to the MGC sample; 
  b) the data were taken in the $r$ band, 
  $\langle\mu_\mathrm{e,b}\rangle=20.48$~mag\,arcsec$^{-2}$ 
  corresponds to the \citet{sim+2011} subsample. In order to show the lack of trend of $\mu_\mathrm{e,b}$ with $n$ for the \citet{sim+2011} subsample, the values of $\mu_\mathrm{e,b}$ were avaraged inside the bin $\Delta n=0.5$. The corresponding bars represent the standard deviation of $\mu_\mathrm{e,b}$ inside each bin.
  The samples used are listed in Table~2.
}
\label{muebn}
\end{figure}

To sort out these questions, let us first consider the 
relation $\mu_\mathrm{e,b}$ vs $n$. Surprisingly, being primarily 
measured (as for ellipticals) or fitted (as for bulges), 
$\mu_\mathrm{e,b}$ shows no trend with $n$. The top plot in 
Fig.~\ref{muebn} demonstrates data from several samples of 
elliptical galaxies and of spiral galaxy bulges in the $B$ band. 
The compiled sample is inhomogeneous; the scatter is large. 
Some points fall off the main distribution. This is a case 
of bright galaxies by \citet{mol2004}. But both for the entire sample and for 
each subsample we can not observe the trend. The lack of the trend 
clearly manifests itself for the largest sample of galaxies from MGC. 
The sample is poorly inhabited in the region of large values of 
$n$, but the general behaviour is unambiguous. The straight line 
shows the median value of $\mu_\mathrm{e,b}$ for MGC's galaxies. 
Bright galaxies from \citet{mol2004} are above this line contributing only 
to the scatter, but without creating the trend.

The most impressive example is shown in Fig.~\ref{muebn}b. 
It shows the data in the $r$ band for the bulges from the 
sample by \citet{sim+2011}. The data were complemented by several 
additional samples with available decompositions in the same $r$ band 
as in Fig.~\ref{mu0blgn}b. In Fig.~\ref{muebn}b the data from the very 
homogeneous sample of spiral galaxies by G09 are also 
plotted. It is clearly seen that the points merely scatter around the 
median value of $\mu_\mathrm{e,b}$ 
(straight line\footnote{The median value of  
$\mu_\mathrm{e,b}$ was calculated for galaxies from the sample by 
\citet{sim+2011}.}). There is no trend of $\mu_\mathrm{e,b}$ with $n$. 
There is a slight bend around the median value of $\mu_\mathrm{e,b}$ 
for Simard's (\citeyear{sim+2011}) data. The procedure 
of $\mu_\mathrm{e,b}$ deriving is not direct for this sample, and the 
fitting procedure itself can be a reason of existence of that 
bend\footnote{For this sample the free fitting parameters were the 
total flux, the bulge fraction $B/T$, the effective radius 
$r_\mathrm{e,b}$, and the \ser\ index $n$. 
The values of $\mu_\mathrm{e,b}$ and $\mu_\mathrm{0,b}$ should be 
calculated through these parameters via appropriate formulas.}.
Faint Virgo cluster galaxies \citep{mcdonald+2011} and bright galaxies 
from \citet{mol2004} deviate from the straight line lying above and 
below the median line.

The lack of the trend of $\mu_\mathrm{e,b}$ with $n$ was neither 
noted nor discussed earlier but helps us to understand the relation 
between $\mu_\mathrm{0}$ and $n$. As $\nu_\mathrm{n}$ is an almost linear function of $n$ (see Eq.~\ref{bn}), $\mu_\mathrm{0,b}$ can be expressed as:
\begin{equation}
\mu_\mathrm{0,b} \sim 
\langle\mu_\mathrm{e,b}\rangle - 1.0857 \, (2n-0.33) \, ,
\label{form_mu0_mue}
\end{equation}
where $\langle\mu_\mathrm{e,b}\rangle$ is the median value for a 
sample. This is a linear self-relation between $\mu_\mathrm{0,b}$ and $n$.

As $\nu_n$ is an almost linear function of $n$ (see Eq.~(\ref{bn})), 
we have a linear self-relation between $\mu_\mathrm{0,b}$ and $n$  (see Fig.~20 for the sample by G09) that 
transforms into a curved self-relation between $\mu_\mathrm{0,b}$ 
and $\log (n)$ (see Fig.~\ref{mu0blgn} and  Fig.~\ref{m0b_lgn_grah}).

\subsubsection[]{The relation between central surface brightness 
and \ser\ index: the Gadotti's sample}
\label{ss_m0bn}

\begin{figure*}
\centering
\includegraphics[width=18.0cm, angle=0]{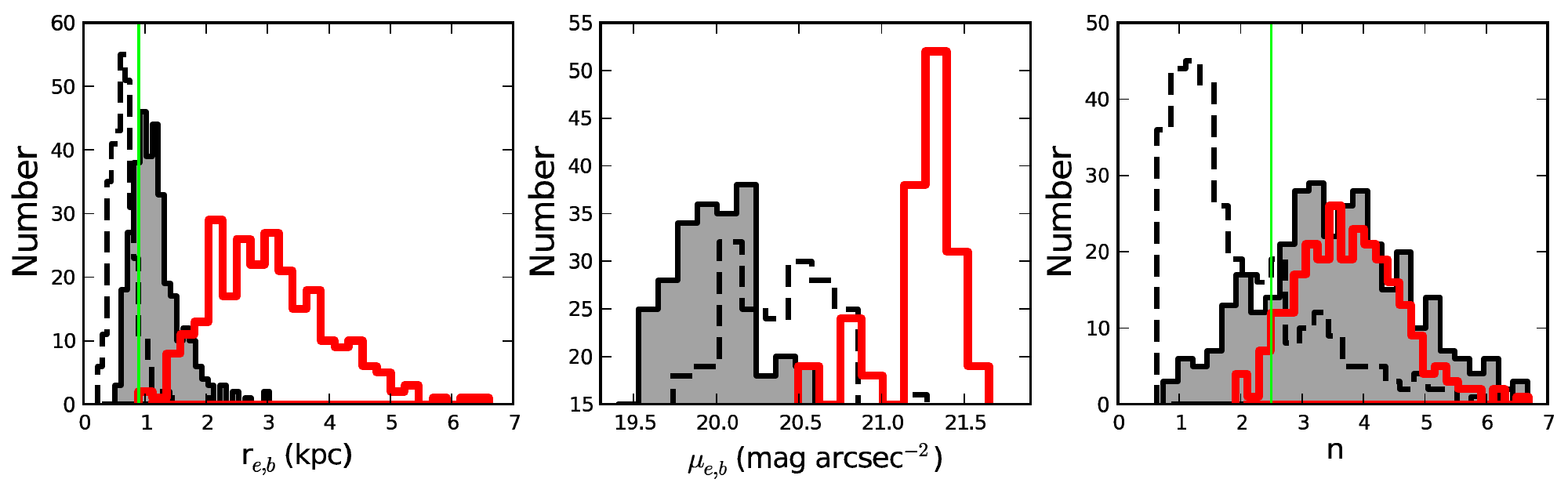}
\caption{Distributions over parameters 
  $r_\mathrm{e,b}$, $\mu_\mathrm{e,b}$, and $n$ for the G09 sample. The division into three separate subsamples as it is 
  found in G09. The black dashed line corresponds to faint bulges, 
  the gray filled histogram is plotted for bright bulges, and the 
  red solid line corresponds to elliptical galaxies. Vertical lines represent values of $r_\mathrm{e,b} = 0.9$~kpc and $n = 2.5$ to discriminate subsamples with bright and faint bulges.}
\label{gadhist}
\end{figure*}

To prove the above conclusion, we analyzed carefully the 
fiducial sample by G09 (we will often address to this sample further and use the data from the $r$ band). 
This sample contains a large amount of objects which were 
selected and decomposed very carefully. Thus, the data can be 
considered as quite  homogeneous.
Here we present the distributions over fitted photometric parameters 
$r_\mathrm{e,b}$, $\mu_\mathrm{e,b}$, and 
$n$ for this sample, both for bulges and ellipticals. We divided the 
bulges into two subsamples 
(faint bulges with $M_\mathrm{bulge} \geq -19.3$ and 
bright bulges with $M_\mathrm{bulge} < -19.3$) 
and considered separetely elliptical galaxies for which the 
bulge-to-total ratio $B/T=1$. 
The distributions are shown in Fig.~\ref{gadhist}. 

The reasons for the division into subsamples were as 
follows. G09 showed that classical bulges ($n \ga 2$), pseudobulges 
($n\la2$), and bright elliptical galaxies are separate groups of 
objects. The most significant parameter separating these objects, is 
the \ser\ index $n$, and we can use boundary values of \ser\ model 
parameters distribution for several populations of objects.
For the G09 sample there are two well visible peakes in distributions 
of $r_\mathrm{e,b}$ and $n$ (see Fig.~\ref{gadhist}). 
At the same time, the effective surface brightness distributions are similar for pseudobulges and classical bulges, and we can not distinguish the peaks of both distributions. 
We took the boundary values $r_\mathrm{e,b} \approx 0.9$~kpc, 
$\mu_\mathrm{e,b} \approx 20.2$~mag\,arcsec$^{-2}$ 
(median value for the subsample of bulges and pseudobulges), and 
$n \approx 2.5$. Then we put these values into the relation (see, 
for example, \citealp{caon+1994,grah1997,grah2005})
\begin{equation}
\begin{array}{rl}
M_\mathrm{bulge} = \mu_\mathrm{e,b} & - 
2.5 \log (n e^{\nu_n} \Gamma(2\,n) / \nu_n^{2\,n}) - \\
& \\
& - 2.5\,\log(2\pi r_\mathrm{e,b}^2) - 36.57 \, ,\\
\end{array}
\label{M_bulge}
\end{equation}
As a result, we received $M_\mathrm{sep}\approx-19.3$~mag for the G09 sample.

\begin{figure*}
\centering
\includegraphics[width=18cm, angle=0, clip=]{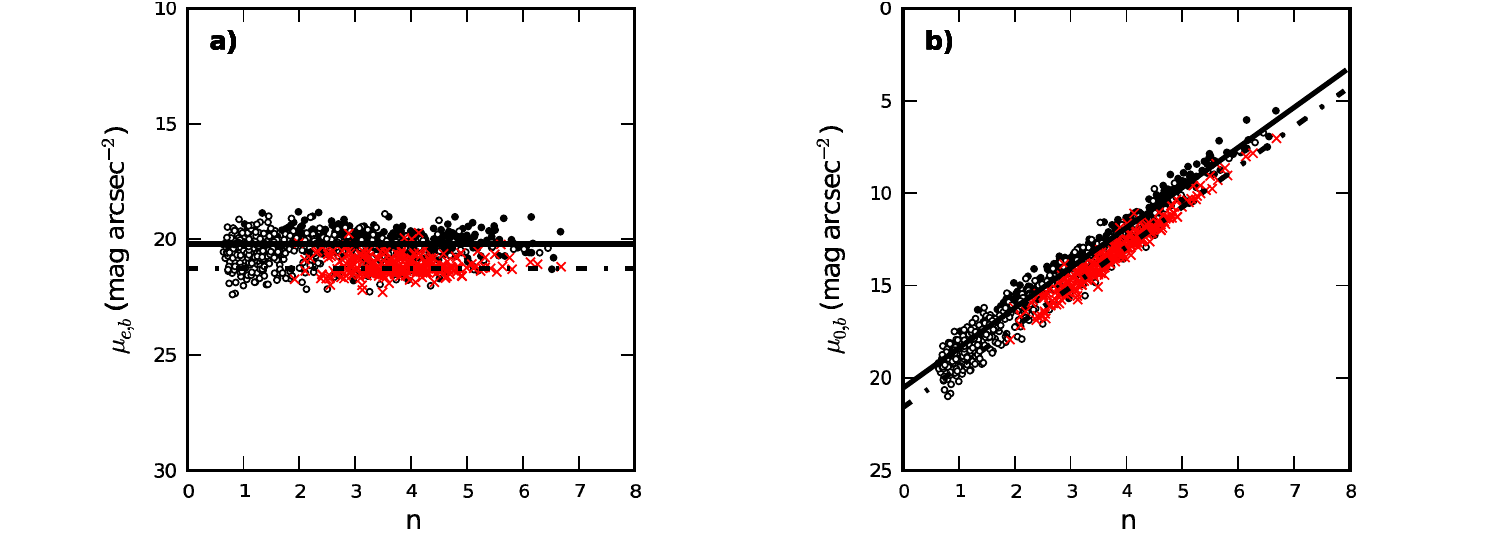}
\caption{
a) The dependence between the effective surface brightness 
$\mu_\mathrm{e,b}$ of the underlying host galaxy bulge or of the 
elliptical and the \ser\ index $n$; 
b) the linear relation between the central surface brightness 
$\mu_\mathrm{0,b}$ and the \ser\ index $n$. 
The data were taken from G09 ($r$ band). 
Black filled circles represent bright bulges 
($M_\mathrm{bulge}<-19.3$~mag), 
black open circles represent faint bulges 
($M_\mathrm{bulge}\geq-19.3$~mag), 
and red crosses correspond to elliptical galaxies. 
The solid line corresponds to the median value 
$\mu_\mathrm{e,b} \approx 20.2$~mag\,arcsec$^{-2}$ 
for the subsample of bulges and pseudobulges, the dash-dotted 
line corresponds to the median value 
$\mu_\mathrm{e,b} \approx 21.25$~mag\,arcsec$^{-2}$ 
for the subsample of ellipticals.} 
\label{gadmu0bn}
\end{figure*}

Fig.~\ref{gadhist} demonstrates three distinct populations of objects. 
The middle plot in 
Fig.~\ref{gadhist} shows the distributions over $\mu_\mathrm{e,b}$. The overall range of 
$\mu_\mathrm{e,b}$ is rather small, no greater than 3~mag, but for 
each subsample the scatter is much smaller (about 1~mag). 
Thus, the distribution of $\mu_\mathrm{e,b}$ gives only the scatter 
around the relation~(\ref{form_mu0_mue}) (see Fig.~\ref{gadmu0bn} which will be discussed below). 

For small samples, the scatter around the relation between 
$\mu_\mathrm{0,b}$ and $n$ is small because of the limited range of 
$\mu_\mathrm{e,b}$ (as for subsamples of faint and bright bulges). 
The wider the distribution over $\mu_\mathrm{e,b}$ and the more 
inhomogeneous the compiled sample, the thicker the lane surrounding 
the relation~(\ref{form_mu0_mue}) is, but the relation itself does 
not ``sink'' in the scatter.

In summary, {\it there is no linear correlation between 
$\mu_\mathrm{0,b}$ and $\log (n)$. There is just an 
equality~(\ref{form_mueb_mu0b}) which reflects the structure of 
the \ser\ model.} The limited range of $\mu_\mathrm{e,b}$ for any 
sample transforms this equality into the linear pseudo-relation 
between $\mu_\mathrm{0,b}$ and $n$ (see Eq.~(\ref{form_mu0_mue})) 
creating a false illusion of a correlation, i.e a self-correlation. 
Moreover, at the limited range of $n$ any linear relation can be 
presented as logarithmic, i.e. depending on $\log (n)$. That is why 
there is no mystery in the widely discussed relation 
$\mu_\mathrm{0,b}$ vs $\log (n)$ 
\citep{bingg1998,kho+2000a,kho+2000b,mol2001,grah2003,aguerri+2004,
ravikumar+2006,barway+2009,kourkchi+2012}. 
The relation is simply the result of a fitting procedure and is 
based on the formula~(\ref{SBr}) for the \ser\ surface brightness 
profile. 

A self-correlation between $\mu_\mathrm{0,b}$ and $n$ follows from 
the fact that $\mu_\mathrm{e,b}$ is independent on $n$ that is well shown for G09 sample in Fig.~\ref{gadmu0bn}a. Bulges, pseudobulges, and elliptical galaxies do not show any trend between $\mu_\mathrm{e,b}$ and $n$. 
Such an independence transforms into the linear pseudo-relation 
(Fig.~\ref{gadmu0bn}b) between $\mu_\mathrm{0,b}$ 
and $n$ with a scatter that reflects the range of $\mu_\mathrm{e,b}$ 
in the samples under discussion. If we use $\log (n)$ instead of $n$, we 
obtain a curved pseudo-relation (Fig.~\ref{mu0blgn}). 
The nature of the curvature in Fig.~\ref{PhP1} is exactly the same. 
The PhP1 includes $\mu_\mathrm{0,b}$ which according 
to~(\ref{form_mueb_mu0b}) comprises $n$. We can approximate 
$n \sim \log (n)$ in a limited range of $n$ and obtain a nearly flat 
photometric plane in the form~(\ref{form_PhP1}). 
In the wider range of $n$ it is curved (Fig.~\ref{mu0blgn}) because 
the relation between $\mu_\mathrm{0,b}$ and $n$ is linear 
(Fig.~\ref{gadmu0bn}). In the next section, we show that the parameter $r_\mathrm{e,b}$ involved in the 
relation~(\ref{form_PhP1}) does not affect our conclusion.

\subsubsection[]{Effective radius vs \ser\ index for bulges 
and ellipticals}

The existence of the univariate correlation between $r_\mathrm{e,b}$ 
and $n$ that might diminish the scatter in the bivariate relation, 
is very doubtful. Some authors revealed a correlation between 
$r_\mathrm{e,b}$ and $n$ 
\citep{caon+1993,grah+1996,gutierrez+2004,mol2004,labarbera+2004,labarbera+2005}. 
\cite{kho+2000b} and \cite{mol2001} give the linear correlation 
coefficient for this correlation to be $\rho > 0.6$ with a 
significance level of 99.98~\%. 
\cite{men+2008} were less enthusiastic about this correlation and 
estimated $\rho=0.28$ for their sample of S0-Sb galaxies in the $J$ 
band. \cite{aguerri+2004} analyzed the photometry of 
116 bright galaxies from the Coma cluster and found the relation 
between $r_\mathrm{e,b}$ and $n$ to be statistically insignificant 
($\rho=0.46$, $P=0.07$).
\cite{ravikumar+2006} demonstrated that a plot of the \ser\ index 
against the effective radius shows the presence of two broad 
distributions (for Es and bright bulges, for dEs and faint bulges 
of S0s and spirals), but without a good correlation within each group. 
\cite{barway+2009} found systematic differences between bright 
and faint lenticulars with respect to the \ser\ index as a function 
of the effective radius. Bright lenticulars are well correlated 
($\rho=0.79$ with significance greater than 99.99~\%), but faint 
lenticulars do not show any correlation. 

G09 sorted out the question about the correlation between 
$r_\mathrm{e,b}$ and $n$. He demonstrated that systems with larger 
$n$ tend to be more extended but this tendency is rather weak. 
The \ser\ index $n$ does slowly rise with $r_\mathrm{e,b}$ for bulges, 
but it is rather constant for ellipticals. 

\begin{figure}
\centering
\includegraphics[width=3.0in, angle=0]{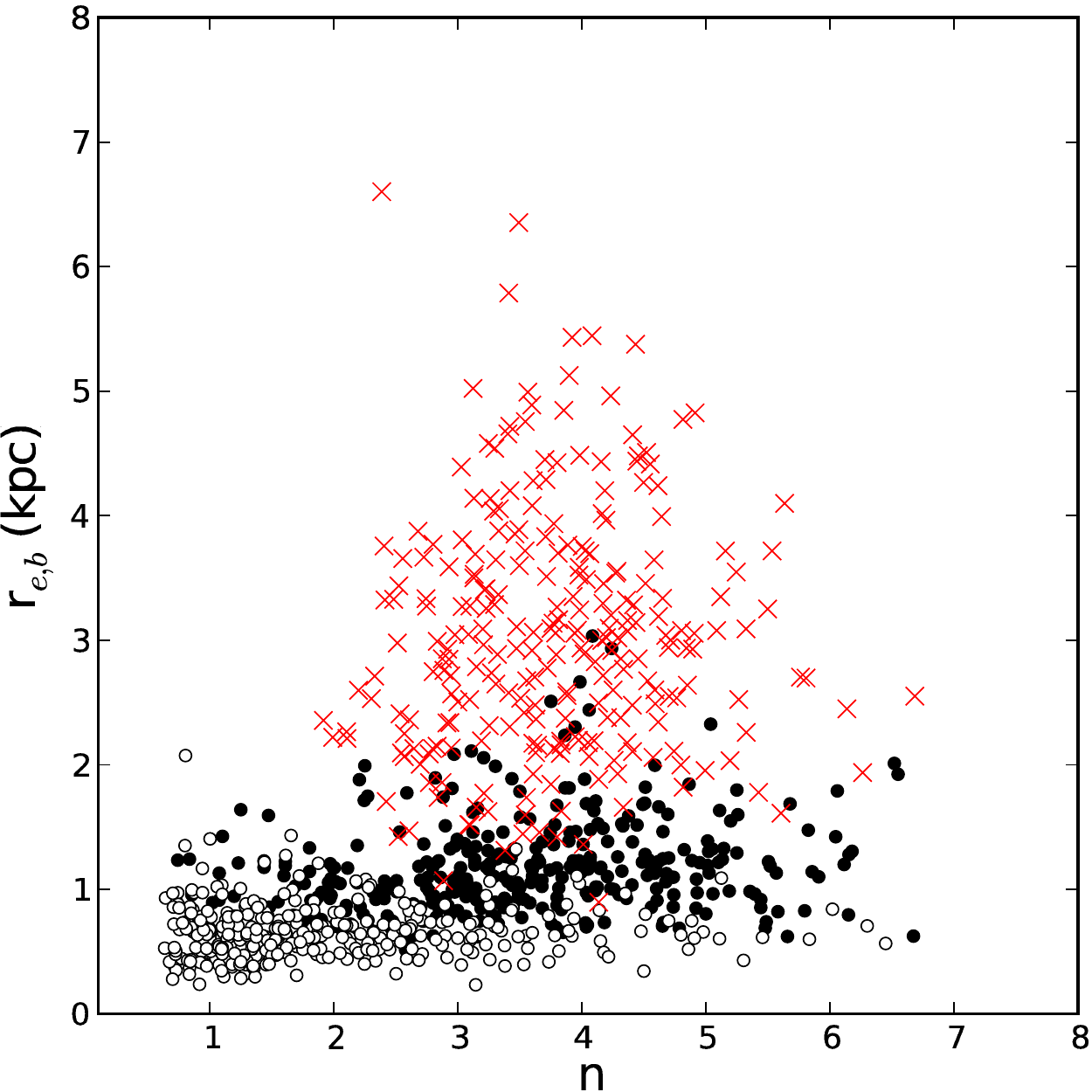}
\caption{The dependence between the bulge effective radius and 
  the \ser\ index plotted for the sample by G09 in the $r$ band
  (symbols as in Fig.~\ref{gadmu0bn}).}
\label{gadcors}
\end{figure}

We replotted these correlations for the sample by 
G09 in the $r$ band. Fig.~\ref{gadcors} represents the plane 
$r_\mathrm{e,b}$~--~$n$. 
One can see that objects from three groups (faint bulges, bright 
bulges, and ellipticals) occupy quite different areas in 
Fig.~\ref{gadcors}. Bulges and elliptical galaxies are almost 
perpendicular to each other. 

For bulges (rather faint objects) the scatter of $r_\mathrm{e,b}$ 
is small. On the contrary, the range of $n$ is large. Thus, the 
effective radius is almost independent on $n$. At the same time, 
bright galaxies (ellipticals) barely show the trend of 
$r_\mathrm{e,b}$ with $n$ along the wide area that is almost 
perpendicular to the $n$ axis. An inhomogeneous 
sample containing the random mixture of bulges and ellipticals, 
faint and bright objects can produce the false correlation 
between $r_\mathrm{e,b}$ and $n$. 

\begin{figure}
\centering
\includegraphics[width=3.5in, angle=0, clip=]{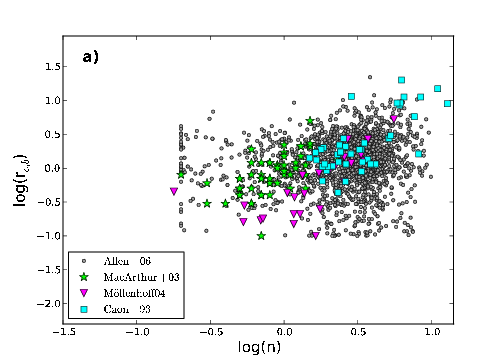}
\includegraphics[width=3.5in, angle=0, clip=]{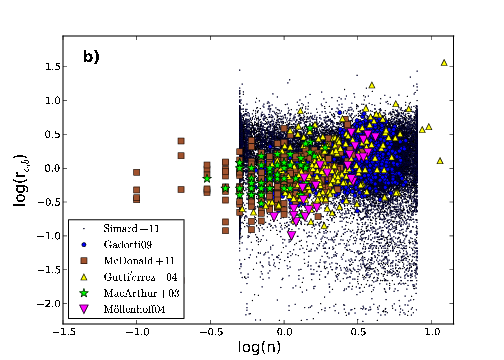}
\caption{The effective radius versus the \ser\ index for 
  a) the $B$ band; 
  b) the $r$ band.
  The samples used are listed in Table~2.}
\label{re_n}
\end{figure}

Additionally, we have gathered data from the samples listed in 
Table~2 and plotted $r_\mathrm{e,b}$ against $n$ in the $B$ and $r$ 
bands (Fig.~\ref{re_n}). The result is very convincing. One can 
notice a slight trend only for small samples, but the overall picture 
does not show any correlation.

Here, it should be noted that we might be missing a significant part of galaxies with small bulges. The selection effects may put a low limit on the distribution of $r_\mathrm{e,b}$ which may change the view of found (or unfound) correlations. However, in this paper we do not consider the inhomogeneity or the completness of the samples.

Thus, the effective radius is not a third parameter that can improve 
the relation~(\ref{form_PhP1}). 
Moreover, in the relation~(\ref{form_PhP1}) $r_\mathrm{e,b}$ is 
expressed in kpc, so the term with $\log(r_\mathrm{e,b})$ can give a 
very small contribution in the $x$-axis expression presented in 
Fig.~\ref{PhP1}, and the leading relation in the 
expression~(\ref{form_PhP1}) is a 
self-correlation~(\ref{form_mu0_mue}).

\subsection[]{Photometric Plane 1 as a self-relation}
\label{ss_PhP1}

\begin{figure*}
\centering
\includegraphics[width=18cm]{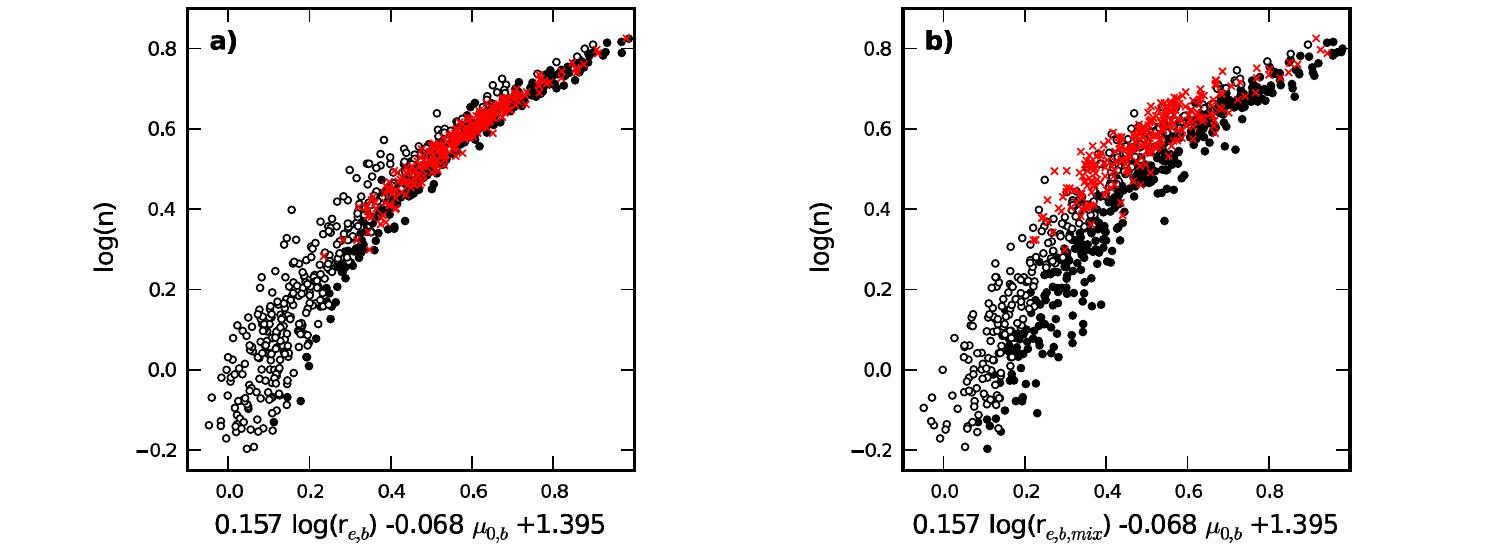}
\caption{The Photometric Plane (PhP1) for the G09 sample in the $r$ 
  band; 
  a) real data;
  b) G09 sample but with randomly mixed values of $r_\mathrm{e,b}$  
  (symbols as in Fig.~\ref{gadmu0bn}).}
\label{PhP1Gad}
\end{figure*}

To demonstrate the insignificance of the contribution of the term 
with $r_\mathrm{e,b}$ which inputs only noise in the 
relation~(\ref{form_PhP1Gad}), we constructed the PhP1 for the G09 
sample (Fig.~\ref{PhP1Gad}a). For the overall sample 
we fitted the expression for the PhP1 as:
\begin{equation}
\log (n) = 0.157 \log (r_\mathrm{e,b}) - 0.068 \, \mu_\mathrm{0,b} + 1.395 
\, .
\label{form_PhP1Gad}
\end{equation}
In the relation~(\ref{form_PhP1Gad}) the contribution of the term with 
$r_\mathrm{e,b}$ even for large galaxies 
($\log (r_\mathrm{e,b}) \approx 0.5$) is smaller than the scatter due 
to $\Delta \mu_\mathrm{e,b} \approx 3^m$ (see Fig.~\ref{gadhist}). 
Using (\ref{form_mu0_mue}), we rewrote the relation 
(\ref{form_PhP1Gad}) in the form
\begin{equation}
\log (n) = 0.157 \log \langle r_\mathrm{e,b}\rangle - 
0.068 \langle\mu_\mathrm{e,b}\rangle + 0.074 \, \nu_n + 1.11 
\, ,
\label{form_PhP1Gadmed}
\end{equation}
where $\langle r_\mathrm{e,b}\rangle$ and 
$\langle\mu_\mathrm{e,b}\rangle$ are the median values of the 
effective radius and of the effective surface brightness, 
respectively. Now one can see that the PhP1 is simply another 
representation of the self-correlation~(\ref{form_mueb_mu0b}), 
and the curvature of the PhP1 just shows the curvature of the 
expression for $\mu_\mathrm{0,b}$ via $\log (n)$. 

The following trick helps to prove our main idea. 
We replotted the PhP1 (Fig.~\ref{PhP1Gad}b) with the same 
expression~(\ref{form_PhP1Gad}) but mixed $r_\mathrm{e,b}$ values.
As it has been done earlier, we split up the sample into three groups of galaxies: faint bulges (with $M_r \geq - 19.3$), 
bright bulges (with $M_r < - 19.3$), and ellipticals.
The scatter in Fig.~\ref{PhP1Gad}b increased, mainly for bright and 
large galaxies with a wide range of $r_\mathrm{e,b}$, but the overall 
shape of the dependence did not change. The curvature of the relation 
remains the same because the leading and trivial relation (self-correlation) 
between $\mu_\mathrm{0,b}$ and $\log (n)$ is curved.

\cite{ravikumar+2006} and \cite{barway+2009} noticed that different 
objects (ellipticals, bulges, faint and bright galaxies) form 
different photometric planes with different thicknesses. They noted 
that ellipticals and bulges of bright lenticulars have a tight 
Photometric Plane (PhP1), and they connected this fact with processes 
that lead to relaxed objects. Now we can see that the main difference 
in the PhP1s for samples used in our analysis, is the difference in 
the median value of $\mu_\mathrm{e,b}$ which shifts the plane. 
The range of $\mu_\mathrm{e,b}$ defines the thickness of the plane.

Thus, \textit{there is no mystery in the existence of the PhP1 which 
is simply a self-correlation contaminated by the term 
$r_\mathrm{e,b}$.} 

In Fig.~\ref{PhP1} the coefficient under the term $\mu_\mathrm{0,b}$ 
is about the same (0.064 -- 0.068) in different bands while the 
coefficient under the term $r_\mathrm{e,b}$ varies substantially. 
This proves that the leading relation in the 
expression~(\ref{form_PhP1}) is the linear dependence of 
$\mu_\mathrm{0,b}$ on $n$ with the proviso that 
$\mu_\mathrm{e,b}$ is independent on $n$. But two intriguing questions remain. 
Why is the range of $\mu_\mathrm{e,b}$ for different objects rather 
small (on average, not greater than $5^m$ while the luminosity can 
change up to 8--9 magnitudes), and why $\mu_\mathrm{e,b}$ does not 
correlate with $n$?

\subsection[]{Photometric Plane 2}
\label{ss_PhP2}

Using a sample of early-type galaxies from the Virgo and Fornax 
clusters with photometric parameters in the $B$ band, \cite{grah2002} 
constructed the Photometric Plane as a variant of the Fundamental 
Plane \citep{djorg1987,dressler+1987} in which the \ser\ index $n$ 
has replaced the central velocity dispersion
\begin{equation}
\log (r_\mathrm{e,b}) = 
a \log (n) + b \langle\mu\rangle_\mathrm{e,b} + c \, ,
\label{form_PhP2}
\end{equation}
where $\langle\mu\rangle_\mathrm{e,b}$ is the mean surface brightness 
within the effective half-light radius. The value of 
$\langle\mu\rangle_\mathrm{e,b}$ can be obtained from 
$\mu_\mathrm{e,b}$ following, for example, 
\citet{caon+1994} and \cite{grah1997}:
\begin{equation}
\langle\mu\rangle_\mathrm{e,b} = \mu_\mathrm{e,b} - 
2.5 \log (n e^{\nu_n} \Gamma(2\,n) / \nu_n^{2\,n}) \, .
\label{mu_mean_eff}
\end{equation}

\cite{grah2002} was motivated by the fact that $n$ quantifies the 
degree of mass concentration of a galaxy, and the central velocity 
dispersion traces the mass of a galaxy. Both these quantities may 
correlate. \cite{grah2002} found such a correlation and introduced 
the Photometric Plane (PhP2 in our notation) in the form given by 
the expression~(\ref{form_PhP2}). 
\cite{labarbera+2004,labarbera+2005} derived the similar Photometric 
Plane in the $K$ band for early-type galaxies in two rich clusters at 
$z \sim 2$ and $z \sim 3$. 

Without the term $\log (n)$ the relation~(\ref{form_PhP2}) is simply a 
version of the Kormendy relation which is a univariate relation 
between $\log (r_\mathrm{e,b})$ and $\langle\mu\rangle_\mathrm{e,b}$. 
If there is another univariate relation between 
$\log (r_\mathrm{e,b})$ and $\log (n)$, we can reduce the scatter by 
constructing the bivariate relation~(\ref{form_PhP2}). 
\cite{grah2002} claimed that the scatter in $\log (r_\mathrm{e,b})$ 
about the $\log (r_\mathrm{e,b})$~--~$\log (n)$ relation for the data 
analysed was 0.35 dex while the scatter for the Kormendy 
relation between 
$\log (r_\mathrm{e,b})$ and $\langle\mu\rangle_\mathrm{e,b}$ 
was 0.25 dex. 
Using all three photometric parameters resulted in a tighter 
correlation with the scatter of 0.125 dex,
\cite{labarbera+2004,labarbera+2005} came to the similar conclusion.

\begin{figure*}
\centering
\includegraphics[width=18cm, angle=0, clip=]{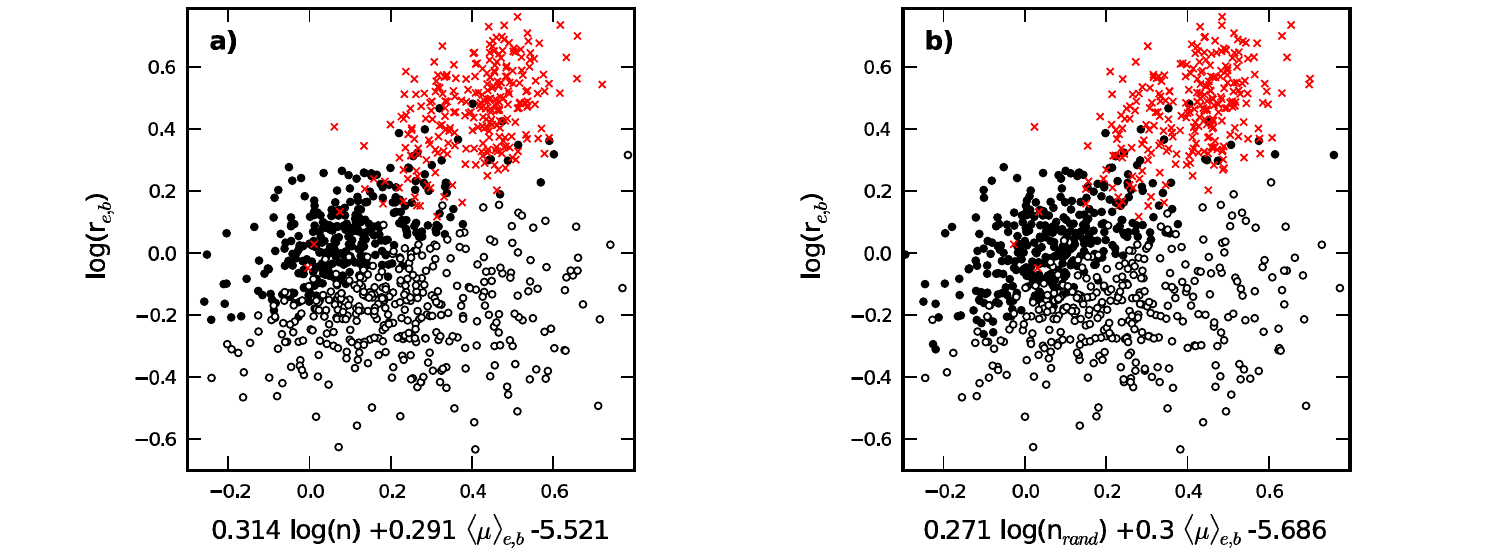}
\caption{The Photometric Plane (PhP2) built for the sample by G09 
  in the $r$ band
  with true (a), i.e. taken from original papers, and random 
  (b) values of the S\'ersic index $n$. 
  The limits of randomly taken $n$ were from 0.5 to 8 
  (symbols as in Fig.~\ref{gadmu0bn}).}
\label{PhP2art}
\end{figure*}

To check the above conclusion, we constructed the PhP2 in the 
form~(\ref{form_PhP2}) for the G09 sample 
(Fig.~\ref{PhP2art}a). Again, we distinguished between faint and bright bulges, and elliptical galaxies. We constructed PhP2 only for 
bright bulges and then superimposed 
on this plane all other galaxies from the G09 sample.

The scatter of points in Fig.~\ref{PhP2art}a is very big. 
The relation is seen only for rather bright ellipticals and bulges 
(black filled circles). 

First, this scatter is due to the lack of the correlation 
between $\log (r_\mathrm{e,b})$ and $\log (n)$ 
(see, for example, Figs.~\ref{gadcors}~and~\ref{re_n}). 
Such a correlation reveals itself only for small samples, especially 
if they contain objects from the bottom left and top right corners as 
in Fig.~\ref{gadcors}, creating the illusion of the correlation. 
Secondly, the sample comprises faint objects that can deviate from 
the leading relation between 
$\log (r_\mathrm{e,b})$ and $\langle\mu\rangle_\mathrm{e,b}$.

To demonstrate the insignificance of the contribution of the term with 
$\log (n)$ in the relation~(\ref{form_PhP2}), we took random values of 
$n$ for the the sample by G09 and reconstructed the PhP2 for bright 
objects (Fig.~\ref{PhP2art}b) to compare it with the original  data
(Fig.~\ref{PhP2art}a). 

The most surprising thing is that the 
coefficient under the term $\langle\mu\rangle_\mathrm{e,b}$ retained 
its value, and it turned out to be robust. The coefficient under the 
term $\log (n)$ was slightly altered, but the overall picture did not 
change. This proves the littleness of the term $\log (n)$ in the 
relation~(\ref{form_PhP2}).

Without the term of $\log (n)$ the relation~(\ref{form_PhP2}) is a 
version of the Kormendy relation in which $\log (n)$ creates noise. 
Bright galaxies form a rather tight relation in Fig.~\ref{PhP2art}a, 
and faint galaxies deviate from the relation in exactly the same 
way as in the Kormendy relation.

\subsection[]{Photometric Plane 3}
\label{ss_PhP3}

\cite{kourkchi+2012} used a sample of dwarf galaxies in the Coma 
cluster in magnitude range $-21 < M_I < - 15$ to construct 
the Photometric Plane in the form
\begin{equation}
\langle\mu\rangle_\mathrm{e,b} = a \log (r_\mathrm{e,b}) + b \log (n) + c \, .
\label{form_PhP3}
\end{equation}
We designate this Photometric Plane as PhP3. The motivation of 
\cite{kourkchi+2012} was the same as in \cite{grah2002}. They wanted 
to simplify the Fundamental Plane. Replacing the velocity dispersion 
with the \ser\ index, the PhP3 is obtained more economically than 
the Fundamental Plane because it is based only on photometric 
parameters. \cite{kourkchi+2012} also claimed that the scatter in 
the relation~(\ref{form_PhP3}) diminished in comparison with two 
appropriate univariate relations.

\begin{figure*}
\centering
\includegraphics[width=18cm, angle=0, clip=]{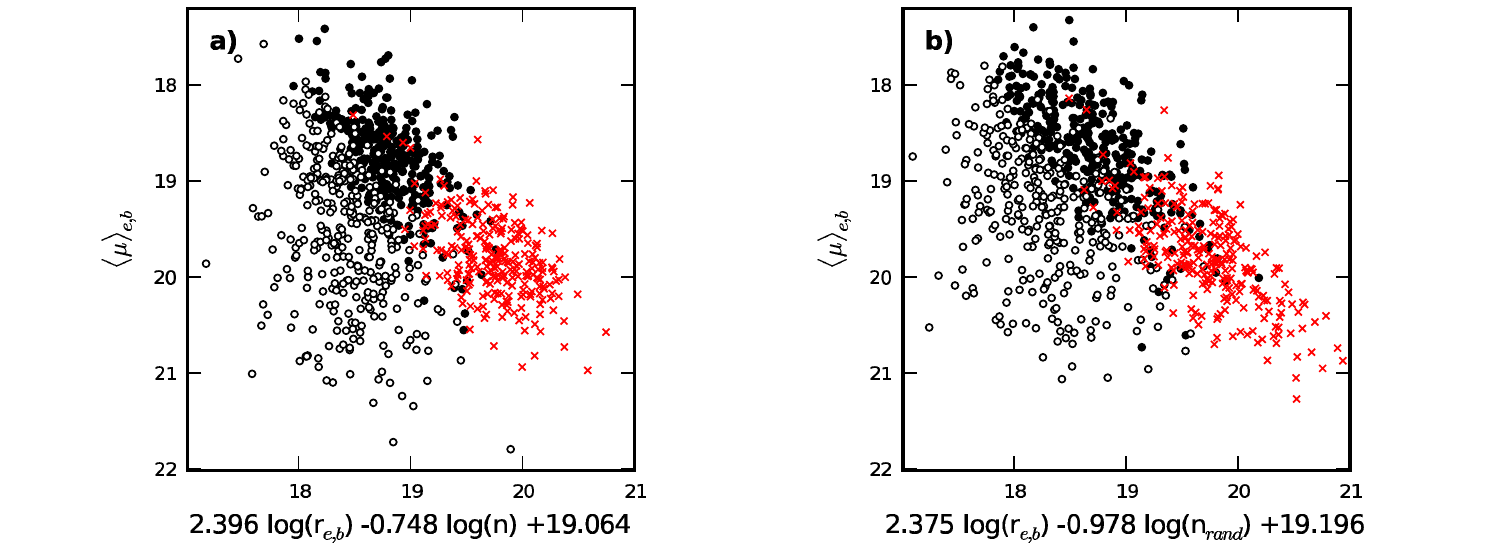}
\caption{The Photometric Plane (PhP3) built for the sample by G09 
  in the $r$ band with true (a), i.e. taken from original papers, 
  and random (b) values of the S\'ersic index $n$ 
  (symbols as in Fig.~\ref{gadmu0bn}). 
  The range of randomly taken $n$ was from 0.5 to 8.}
\label{PhP3art}
\end{figure*}

As before, we constructed the PhP3 in the form~(\ref{form_PhP3}) from the sample 
by G09 (Fig.~\ref{PhP3art}a) with division into
faint bulges, brigh bulges, and ellipticals. 
Objects from these three groups occupy quite 
different areas in Fig.~\ref{PhP3art}a. 
Faint bulges have a larger scatter in the plane than bright 
bulges and ellipticals. The relation~(\ref{form_PhP3}) comprises 
a weak self-correlation between $\langle\mu\rangle_\mathrm{e,b}$ and 
$n$ (see the relation~(\ref{mu_mean_eff})) 
which can affect the bivariate relation~(\ref{form_PhP3}). 

We destroyed this self-correlation by using random values of $n$. 
The reconstructed PhP3s are demonstrated in Fig.~\ref{PhP3art}b. 
We can see that the term $\log (n)$ does not affect the overall shape of 
the relation, and it can be included in an arbitrary form. Thus, the 
relation~(\ref{form_PhP3}) is a worse version of the Kormendy 
relation. It is not surprising that bright and faint bulges follow 
their own relations, and for the bright bulges the relation is 
tighter. The coefficient under the term with $\log (r_\mathrm{e,b})$ is 
about 2.4 which is close to the slope of the Kormendy relation.

\subsection[]{Conclusion}

In Sect.~\ref{s_corr_bulges} we discussed various versions of the 
so-called Photometric Plane. This plane joins photometric 
characteristics 
($n$, $\mu_\mathrm{0,b}$ or $\langle\mu\rangle_\mathrm{e,b}$, and 
$r_\mathrm{e,b}$.) of ellipticals and bulges of spiral galaxies. 

As we have shown, the Photometric Plane has no independent physical 
sense --- it simply reflects $\mu_\mathrm{0,b}$~--~$n$ self-correlation
(PhP1, Sect.~\ref{ss_PhP1}) or is an entangled version of the Kormendy
relation (PhP2 and PhP3, Sect.~\ref{ss_PhP2} and~\ref{ss_PhP3}).

\section[]{Kormendy relation}
\label{s_KR}

In this section we are about to touch on the essence of the above 
noted Kormendy relation which has being widely discussed for several 
decades.

\subsection[]{Graham's approach: outline}

In some works we can find the idea that the Kormendy relation (and 
other similar to it curved relations between photometric parameters 
of ellipticals and of galaxy bulges) can not be considered as the 
evidence of the physical difference between dwarfs and bright 
ellipticals. The same issue stands for galaxies with pseudobulges 
and bright classical bulges. This point of view has been actively 
propagated by Graham in several articles 
\citep[e.g.][]{grah2003,grah2011,grah2013a}. 
His attempt to unify dwarfs and giant galaxies stems from the idea that the curved relations between photometric parameters are the result of the existence of two linear relations valid for both classes of objects. Graham's conclusions were 
extended not only to elliptical galaxies, but also to disc galaxies 
with classical bulges and pseudobulges \citep{grah2013b}. 
In this section we check these conclusions about the curved 
relations operating with the already used G09 sample as the best 
representative for bulges from our set of samples. 

\citet{grah2011,grah2013a} uses two main correlations between the 
\ser\ model parameters. These are the correlation between the 
luminosity and the central surface brightness of a spherical 
component \citep[e.g.][]{bingg1998,grah2003}, and the relation 
between the luminosity and the \ser\ index 
\citep[e.g.][]{caon+1993,young1994,bingg1998,mol2001,grah2003,ferra+2006}. 
In the $B$ band he found:
\begin{equation}
\begin{array}{l}
M_\mathrm{bulge} = 0.67 \mu_\mathrm{0,b} - 29.5 \, , \\
M_\mathrm{bulge} = -9.4 \log(n) - 14.3 \, .\\
\end{array} 
\label{eq:xdef}
\end{equation}
From the above relations, Graham derived the expression between the 
central surface brightness and the \ser\ index:
$$
\mu_\mathrm{0,b} = 22.8 - 14.1 \log (n) \, .
$$

\citet{grah2013a} noted that this relation is roughly applicable 
only for values of $n \geq 1$ as one can see in fig.~7b from 
\citep{grah2013a}. 

\subsection[]{The usage of Graham's approach for the Gadotti's sample}

\begin{figure}
\centering
\includegraphics[width=3in, angle=0]{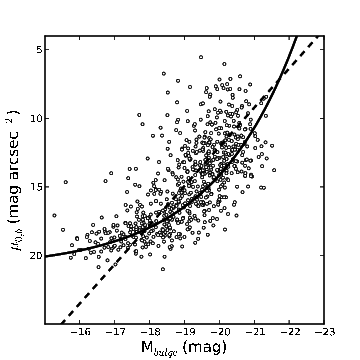}
\caption{The correlation between the central surface brightness and 
  the luminosity of bulges from the G09 sample in the $r$ band. 
  The regression line (dashed line) corresponds to the 
  relation~(\ref{grah2}) while the solid line refers to 
  relations~(\ref{grah22}) and~(\ref{ntrue}) .}
\label{Mb_m0b_grah}
\end{figure}

Let us follow the main idea of \citet{grah2013a} and derive two 
linear relations analogous to~(\ref{eq:xdef}) but for the sample 
by G09. 
The distributions of the main photometric parameters are shown 
in Fig.~\ref{gadhist}. 
The univariate, namely, the relation between 
$\mu_\mathrm{0,b}$ and $M_\mathrm{bulge}$
is reproduced in Fig.~\ref{Mb_m0b_grah} only for bulges. 
We found the regression line (dashed line) for the data. 
It corresponds to the expression
\begin{equation}
M_\mathrm{bulge} = 0.35\,\mu_\mathrm{0,b}  - 24.23 \, .
\label{grah2}
\end{equation}

\begin{figure}
\centering
\includegraphics[width=3in, angle=0]{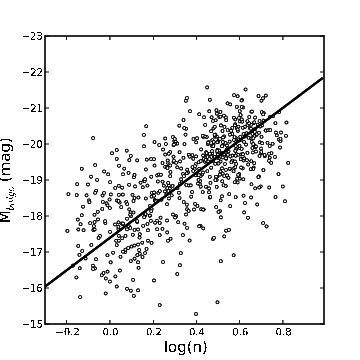}
\caption{The correlation between the luminosity and 
  the \ser\ index (in the logarithmic scale) for bulges from the G09 
  sample in the $r$ band. The regression line corresponds to 
  relation~(\ref{grah22}).}
\label{Mb_n_grah}
\end{figure}

The relation between the \ser\ index and the luminosity
is presented with the regression line (solid line) in 
Fig.~\ref{Mb_n_grah}, also only for bulges.
We expressed the relation as:
\begin{equation}
M_\mathrm{bulge} = -4.51\,\log(n) - 17.394 \, .
\label{grah22}
\end{equation}

From these two relations one can derive:
\begin{equation}
\mu_\mathrm{0,b} = - 12.88\,\log(n) + 19.52\,,
\label{grah1}
\end{equation}

\begin{figure}
\centering
\includegraphics[width=3in, angle=0]{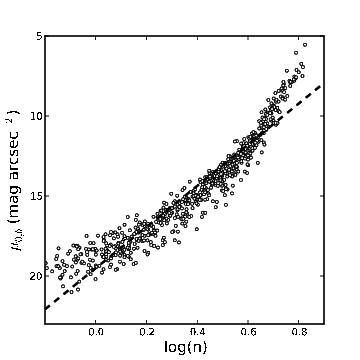}
\caption{The correlation between the central surface brightness 
  and the \ser\ index (in the logarithmic scale) for bulges from the 
  G09 sample in the $r$ band. The regression line corresponds to  
  relation~(\ref{grah1}).}
\label{m0b_lgn_grah}
\end{figure}

The $\mu_\mathrm{0,b}$~--~$n$ relation is presented in 
Fig.~\ref{m0b_lgn_grah} with a regression line corresponding 
to the expression~(\ref{grah1}). Here we forget for a moment about our conclusion 
in Section~\ref{ss_m0bn} that this correlation is linear, but we will 
come back to this fact later.

Then we can repeat the algorithm described in detail in 
\citet{grah2013a}. We take the expression for $M_\mathrm{bulge}$ 
(it follows from \citet{grah2005}) as:
\begin{equation}
M_\mathrm{bulge} = \langle\mu\rangle_\mathrm{e,b} - 
2.5\,\log(2\pi r_\mathrm{e,b}^2) - 36.57\,,
\label{grah3}
\end{equation}
where $\langle\mu\rangle_\mathrm{e,b}$ can be found from 
Eq.~(\ref{mu_mean_eff}).

Now we can eliminate the absolute magnitude from the Eq.~(\ref{grah3}) 
replacing it from relation~(\ref{grah2}) where 
$\mu_\mathrm{0,b}$ can be expressed through 
$\langle\mu\rangle_\mathrm{e,b}$ from~(\ref{mu_mean_eff}) 
and~(\ref{form_mueb_mu0b}). At last, we will have the relation 
between $r_\mathrm{e,b}$ and $\mu_\mathrm{e,b}$ such that
\begin{equation}
\log(r_\mathrm{e,b}) = 0.2\,\left(0.56\,\mu_\mathrm{e,b} - 
A_\mathrm{n} - 12.60 + 0.48\,\nu_{n}\right)\, ,
\label{grahcurved}
\end{equation}
where $A_\mathrm{n}=2.5 \log (n e^{\nu_n} \Gamma(2\,n) / \nu_n^{2\,n})$.

\begin{figure}
\centering
\includegraphics[width=3in, angle=0, clip=]{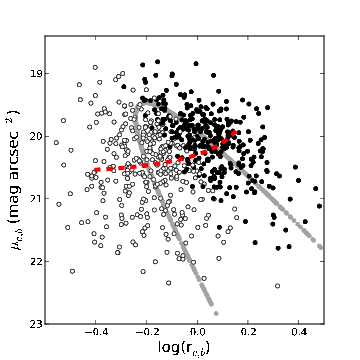}
\caption{The Kormendy relation for the bulges from the G09 sample in 
  the $r$ band with superimposed relations constructed following 
  \citet{grah2013a} and our approach (see the text). 
  Symbols are as in Fig.~\ref{PhP2art}. 
  The line of gray filled circles represents the curved 
  relation~(\ref{grahcurved}), the dashed red line is a 
  relation for which $\mu_\mathrm{e,b}$ is recalculated 
  according to (\ref{ntrue}) instead of (\ref{grah1}) 
 .}
\label{korm}
\end{figure}

Each value of $n$ can be associated with each value of 
$\mu_\mathrm{e,b}$ using relations~(\ref{grah1}) 
and~(\ref{form_mueb_mu0b}). Thus, relation~(\ref{grahcurved}) 
is between $r_\mathrm{e,b}$ and $\mu_\mathrm{e,b}$ only. 
It is curved and shown in Fig.~\ref{korm}. Similar to this algorithm,
we can build other curved relations such as 
$\langle\mu\rangle_\mathrm{e,b}$~--~$r_\mathrm{e,b}$, 
$\mu_\mathrm{e,b}$~--~$M_\mathrm{bulge}$, 
$\langle\mu\rangle_\mathrm{e,b}$~--~$M_\mathrm{bulge}$, 
$r_\mathrm{e,b}$~--~$M_\mathrm{bulge}$ (see \citealp{grah2013a}).

\subsection[]{Our approach for the Gadotti's sample}

\begin{figure}
\centering
\includegraphics[width=3in, angle=0]{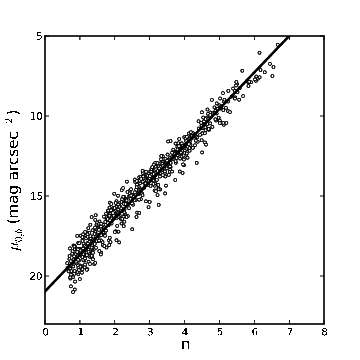}
\caption{The correlation between the central surface brightness of a 
  bulge and the \ser\ index for the G09 sample in the $r$ band. 
  The regression line corresponds to relation~(\ref{ntrue}). Fig.~\ref{mu0blgn} and Fig.~\ref{m0b_lgn_grah} show the same with $\log (n)$.}
\label{m0b_n_grah}
\end{figure}

Let us now turn back to what we showed earlier in 
Section~\ref{ss_m0bn} where the true correlation between the central 
surface brightness and the \ser\ index appeared to be linear. 
For bulges from the sample by G09, this can be written as:
\begin{equation}
\mu_\mathrm{0,b} = -2.28\,n + 20.96\, .
\label{ntrue}
\end{equation}
This relation is presented in Fig.~\ref{m0b_n_grah}. 

Earlier we showed that the effective brightness of elliptical 
galaxies and bulges does not depend on the \ser\ index, and for 
the subsample of bulges by the G09 sample the median value is 
$\langle\mu_\mathrm{e,b}\rangle = 20.21$~mag\,arcsec$^{-2}$. 
Therefore, if we take relation~(\ref{ntrue}) instead 
of~(\ref{grah1}), relation~(\ref{grah22}) presented in 
Fig.~\ref{Mb_n_grah}, and obtain the relation between 
$M_\mathrm{bulge}$ and $\mu_\mathrm{0,b}$, it will be curved. 
We marked this curved relation in Fig.~\ref{Mb_m0b_grah} with 
the solid line. Surprisingly, this curved line represents the data 
better than the linear one and seems to be more appropriate to describe 
$\mu_\mathrm{0,b}$~--~$M_\mathrm{bulge}$ correlation. 
Then, we can repeat the algorithm for building a relation between 
$\mu_\mathrm{e,b}$ and $r_\mathrm{e,b}$ and superimpose it on the data 
in the plane $\mu_\mathrm{e,b}$~--~$r_\mathrm{e,b}$. This relation 
is shown in Fig.~\ref{korm} as a black solid line. We can see that 
this relation does not describe the observed scatter of points at all 
and simply reflects the fact of independence of the parameter 
$\mu_\mathrm{e,b}$ on $n$.

From all of these facts we can conclude that at least for bulges 
presented in the sample by G09, {\it the Graham's approach is not 
applicable in order to explain the Kormendy relation by unifying 
it with two linear relations for faint and bright objects since one 
of them appeared to be wrong.} 


\subsection[]{Another presentation of the Kormendy relation}

Nonetheless, we shall try to apply another approach to check 
whether the Kormendy relation is real and does not follow from any 
unifying relations. To do this, we build several artificial samples which
are described in Table~3. Here we suggest 
the detailed recipe how to construct the Kormendy relation and all 
other correlations for bulges and elliptical galaxies we have already 
discussed. We shall see that such relations have features of the real 
physical difference between galaxies with different photometric 
characteristics and the features of the embedded \ser\ law being 
used to describe a bulge component or an elliptical galaxy. 

We consider distributions of the three \ser\ model 
parameters: 
$r_\mathrm{e,b}$, $\mu_\mathrm{e,b}$, and $n$ from the sample by G09. 
Let us note here that in this subsection we use the whole sample 
including elliptical galaxies. It is done to show more clearly our 
main idea about existence of curved relations (using wider ranges of 
all \ser\ model parameters).
We start with the very simple case of uncorrelated parameters and end 
up with a sample similar to that which was investigated in G09. 

\begin{figure*}
\centering
\includegraphics[height=23.0cm, angle=0]{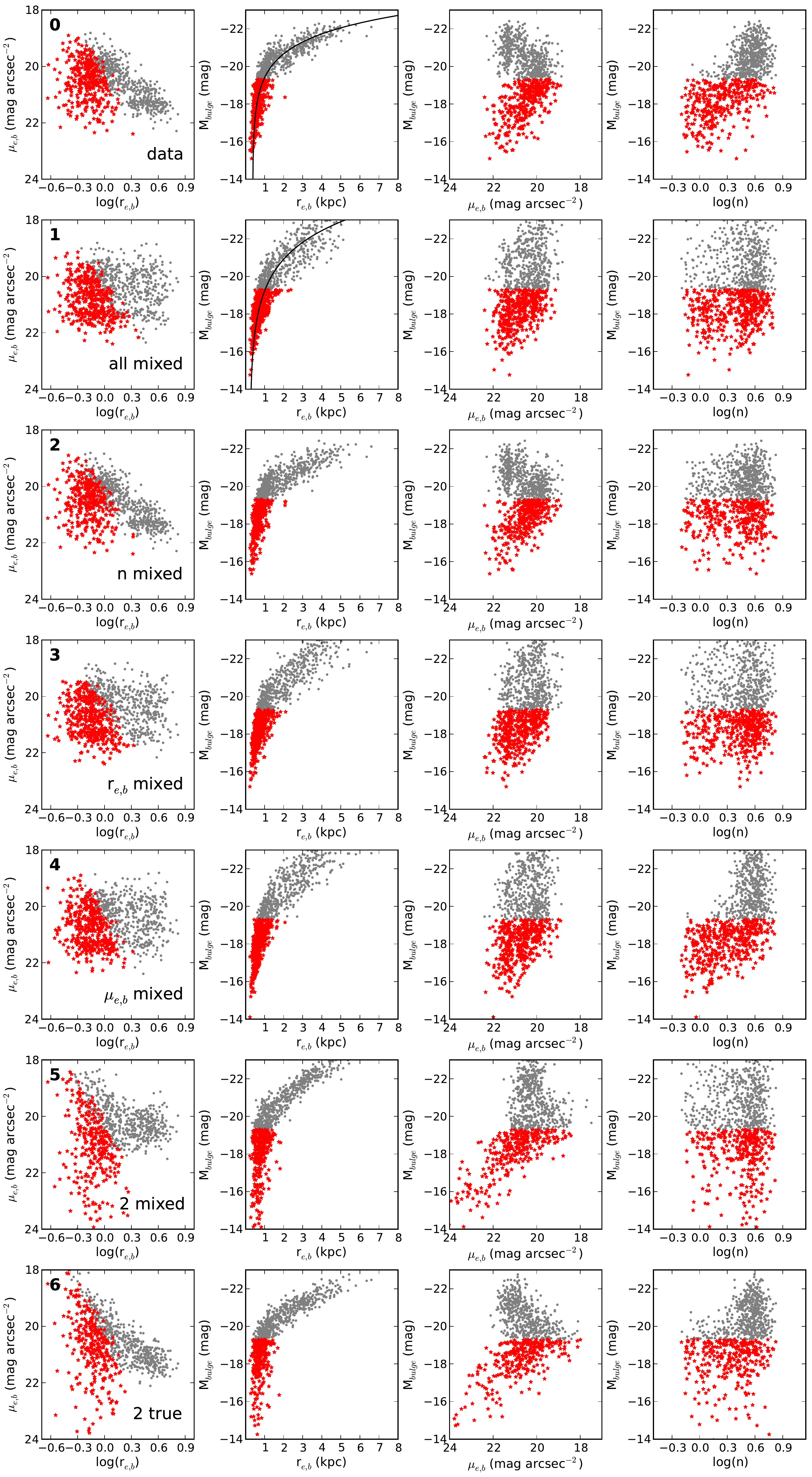}
\caption{Correlations between some photometric parameters 
 plotted for samples from Table~3. Each row represents 
  one sample (row marked as ``0'' represents the distributions for 
  the G09 sample). Red filled stars represent faint bulges 
  ($M_\mathrm{bulge}\geq-19.3$~mag) and grey circles represent bright 
  bulges and elliptical galaxies ($M_\mathrm{bulge}<-19.3$~mag). The solid lines are described in the text.}
\label{simgals}
\end{figure*}

\begin{center}
\begin{table*}
 \centering
 \begin{minipage}{170mm}
  \parbox[t]{170mm} {\caption{List of model samples created to 
   illustrate the appearence of the curved relations. `True' means original data column taken from G09, `mixed' means randomly mixed data column, and `approx' indicate that the parameter was found by using an approximation (see text).}}
  \begin{tabular}{lcccll}
  \hline 
  \hline
Sample \# & $\mu_\mathrm{e,b}$    & $r_\mathrm{e,b}$ & $n$ & Comments & Correlations \\
\hline
0 & true & true & true & G09 sample & All correlations from \citet{gad2009} \\
1 & mixed & mixed & mixed & From G09 & PhP1, $M_b$ vs $r_\mathrm{e,b}$ \\
2 & true & true & mixed & From G09 & PhP1, all Kormendy relations \\
3 & true & mixed & true & From G09 & PhP1, $M_b$ vs $r_\mathrm{e,b}$ \\
4 & mixed & true & true & From G09 & PhP1, $M_b$ vs $r_\mathrm{e,b}$, $M_b$ vs. $n$ \\
5 & approx & mixed & mixed & From G09, $\mu_\mathrm{e,b}$ built from (\ref{grah7}) & PhP1, $M_b$ vs. $r_\mathrm{e,b}$, $M_b$ vs. $n$ \\
6 & approx & true & true & From G09, $\mu_\mathrm{e,b}$ built from (\ref{grah6})  & All correlations from 
\citet{gad2009} \\
 \hline 
 \tabularnewline
 \hline
\end{tabular}
\end{minipage}
  \label{Tab_samples}
\end{table*}
\end{center}
For the comparison, we plotted the correlations for the G09 sample (Fig.~\ref{simgals}(0)) where the luminosity of the spheroidal component $M_\mathrm{bulge}$ was taken from the original tables of the decomposition results.

First, we consider the randomly mixed distribution of values for 
all three parameters from the sample by G09 (see Table~3, sample~\#1). 
It means that we randomly choose the values of this parameters 
from the G09 tables. We calculated the bulge total luminosity $M_\mathrm{bulge}$ via Eq.~(\ref{M_bulge}). 
It is evident that the Kormendy relation and different versions of 
it will not occur because $\mu_\mathrm{e,b}$ and $r_\mathrm{e,b}$ are 
not correlated. We can see in Fig.~\ref{simgals}(1) that there is no 
correlation between the bulge total magnitude $M_\mathrm{bulge}$ and 
$\mu_\mathrm{e,b}$ as well as between $M_\mathrm{bulge}$ and 
the \ser\ index $n$. There is, however, a correlation 
$M_\mathrm{bulge}$ vs $r_\mathrm{e,b}$. The shape of this correlation 
differs from that valid for the sample by G09 (Fig.~\ref{simgals}(0)).

We then build the samples with randomly mixed values of one parameter 
whereas other two parameters may correlate because they are taken 
from the real distributions (see Fig.~\ref{gadhist}). 
It means that we take values of two parameters for given 
galaxies as they are, and the third parameter is chosen randomly 
from the column containing this parameter (samples~\#2--\#4). The luminosity of the spheroidal component $M_\mathrm{bulge}$ for samples~\#2--\#4 was also calculated using Eq.~\ref{M_bulge}.
Some correlations can reveal themselves for such artificial samples, 
but others can be destroyed. For example, the Kormendy relation 
for correlated $r_\mathrm{e,b}$ and  $\mu_\mathrm{e,b}$ (sample~\#2) 
keeps its shape (Fig.~\ref{simgals}(2)), but it is completly pulled 
down for samples~\#3 and~\#4 (Fig.~\ref{simgals}(3),(4)). 
The correlation $M_\mathrm{bulge}$~--~$n$ disappears for 
samples~\#2 and~\#3 but is seen for sample~\#4. 

To create the next sample~\#5, we take the mixed distributions of 
parameters from the $r_\mathrm{e,b}$~--~$n$ plane while values of 
$\mu_\mathrm{e,b}$ are calculated from the Eqs.~(\ref{grah3}) 
and~(\ref{mu_mean_eff}) where $M_\mathrm{bulge}$ is approximated (fitted) for 
sample \#1:
\begin{equation}
M_\mathrm{bulge} = -4.62\,\log(r_\mathrm{e,b} - 0.24) - 19.775 \,.
\label{grah7}
\end{equation}
The line corresponding to this equation is presented in Fig.~\ref{simgals}(1) on the second left plot.
In Fig.~\ref{simgals}(5) one can see that there is something similar 
to the Kormendy relation, but it is quite corrupted while the 
correlation $M_\mathrm{bulge}$ vs $n$ has not yet appeared.

At last, we take the same distribution of parameters from the 
$r_\mathrm{e,b}$~--~$n$ plane as for sample \#4 (i.e. as for the G09 
sample), but $\mu_\mathrm{e,b}$ is calculated from 
the Eq.~(\ref{grah3}) where $M_\mathrm{bulge}$ is found (fitted) from the 
true correlation for the sample by G09 (see the solid line in 
Fig.~\ref{simgals}(0), second left plot):
\begin{equation}
M_\mathrm{bulge} = -3.04\,\log(r_\mathrm{e,b} - 0.38) - 20.03\, .
\label{grah6}
\end{equation}

As we discussed earlier (see Fig.~\ref{gadcors}), the distribution of 
galaxies in the  $r_\mathrm{e,b}$~--~$n$ plane represents two 
perpendicular subsystems: classical bulges + ellipticals and 
pseudobulges. This distribution is actually trimodal 
(see Fig.~\ref{gadhist}) that is very important for understanding 
curved relations as we shall see below.  

In Fig.~\ref{simgals}(6) it is apparent that for the last artificial 
sample~\#6 the built correlations resemble the correlations for the 
G09 sample very well. Therefore, these correlations can be reproduced only when the observed trimodal galaxy distribution in the  $r_\mathrm{e,b}$~--~$n$ plane (see Fig.~\ref{gadcors}) and the observed average scaling relation between $r_\mathrm{e,b}$ and $M_\mathrm{bulge}$ are assumed.

\begin{figure}
\begin{center}
\includegraphics[width=3in, angle=0]{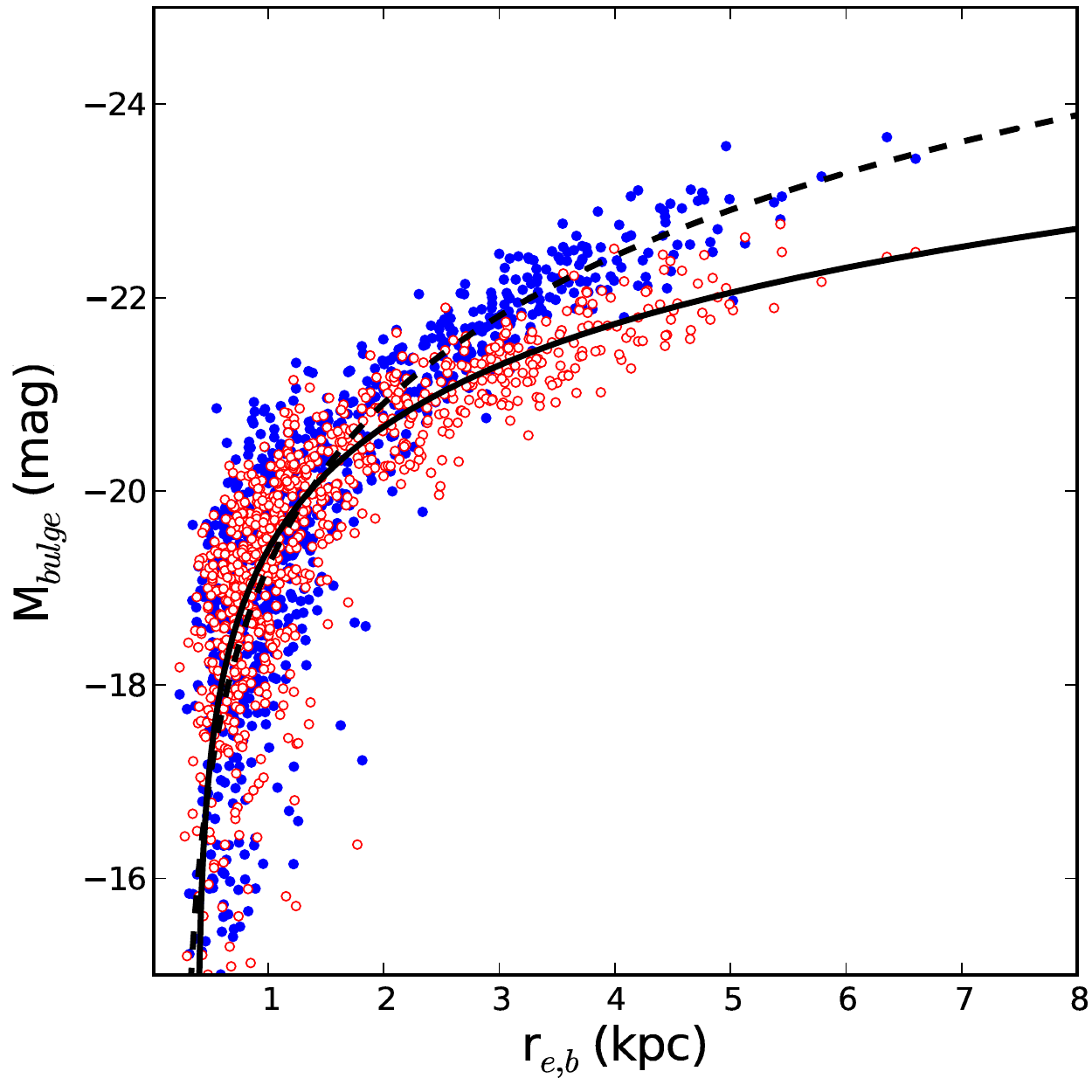}
\caption{Comparison between the two 
  relations~(\ref{grah6}) and~(\ref{grah7}). 
  Red open circles correspond to the data from 
  the G09 sample in the $r$ band, blue filled circles correspond 
  to sample~\#1. 
  Solid and dashed lines represent the approximation of the 
  data for the G09 sample and sample~\#1, respectively 
  (Eqs.~(\ref{grah6}) and~(\ref{grah7})).}
\label{Mbcomp}
\end{center}
\end{figure}

Let us now consider these correlations in more detail. The correlation $M_\mathrm{bulge}$~--~$r_\mathrm{e,b}$ exists for all samples (even for randomly mixed parameter values) because of the presence of the term $r^2_\mathrm{e,b}$ in Eq.~(\ref{grah3}). 
Nevertheless, Fig.~\ref{Mbcomp} shows explicitly that the true distribution of galaxies in the $M_\mathrm{bulge}$~--~$r_\mathrm{e,b}$ plane (i.e. sample \#0) can not be described using only Eq.~(\ref{grah3}). The correlation~(\ref{grah6}) for the last sample~\#6 has a slightly different form than for other samples. In Fig.~\ref{Mbcomp}, one can see 
the difference between the shape of the $M_\mathrm{bulge}$~--~$r_\mathrm{e,b}$ 
correlation for samples~\#5 and ~\#6. The distribution of points in the $M_\mathrm{bulge}$~--~$r_\mathrm{e,b}$ plane for sample~\#6 represents the true variety of galaxies (comparing with sample \#0) what implies that the parameters $r_\mathrm{e,b}$, $\mu_\mathrm{e,b}$, and $n$ are correlated in a certain way. 

The Kormendy relation $\mu_\mathrm{e,b}$ ~--~$r_\mathrm{e,b}$ appeares because of the same fact. The specific form of the correlation  between  $M_\mathrm{bulge}$ and $r_\mathrm{e,b}$ generates the curved relation between  $\mu_\mathrm{e,b}$ and $r_\mathrm{e,b}$. The correlation $M_\mathrm{bulge}$~--~$\mu_\mathrm{e,b}$, in turn, can be retrieved as a consequence of existing correlation $\mu_\mathrm{e,b}$ ~--~$r_\mathrm{e,b}$ as follows from the consideration of sample \#2, where $\mu_\mathrm{e,b}$ and $r_\mathrm{e,b}$ are correlated, and $n$ is a randomly mixed column. Thus, $M_\mathrm{bulge}$~--~$r_\mathrm{e,b}$ relation as well as $M_\mathrm{bulge}$~--~$\mu_\mathrm{e,b}$ relation is another representation of the Kormendy relation.

The correlation $M_\mathrm{bulge}$~--~$n$ is 
a consequence of the bimodality of $n$ and $r_\mathrm{e,b}$. It 
is well seen that bright bulges and elliptical galaxies have little 
scatter of $n$ with the mean value of about 3.4 and the mean 
effective radius larger than that for faint bulges 
(Fig.~\ref{gadhist}). 
The population of faint bulges has a wider distribution of $n$ with 
smaller values of effective radius. Thus, we have two almost 
perpendicular subsystems in the plane $r_\mathrm{e,b}$~--~$n$. 
These two clouds are also presented in the $M_\mathrm{bulge}$~--~$n$ 
plane where bright bulges and elliptical galaxies lie in the upper 
right corner, while faint bulges are in the bottom left corner 
(see Fig.~\ref{simgals}). 
If we had only galaxies with $M_\mathrm{bulge}\geq-19.3$ or 
$M_\mathrm{bulge}<-19.3$, the $M_\mathrm{bulge}$~--~$n$ correlation 
would not be appear! These two separate clouds in the 
$M_\mathrm{bulge}$~--~$n$ plane reveal the correlation between 
these two parameters.

Thus, the curved nature of the considered relations stems from the physical difference between faint and bright bulges. `Unifying' curved relations of the kind of the grey line shown in Fig.~\ref{korm} are the result of the hidden functional correlations of the main fitting parameters $r_\mathrm{e,b}$, $\mu_\mathrm{e,b}$, and $n$, combined with the intrinsic bi- or three modality of their true distributions.





\section[]{Conclusions}
\label{s_conclusions}
We presented the results of the critical review of some widely discussed 
correlations between bulge and disc structural parameters, 
in different photometric bands and of various morphology. The main 
conclusions we can draw from this work can be summarized as follows:

\begin{enumerate}

\item The correlation between the edge-on disc central surface 
brightness reduced to the face-on view and the relative thickness 
of the stellar disc is a self-correlation. It can not empirically 
confirm that faint discs are, on average, more thin than discs of 
higher surface brightness. This correlation only appears as the 
consequence of application of the transparent exponential disc 
model and reflects the distribution characteristics of the sample 
studied.

\item There is no correlation between the effective 
surface brightness $\mu_\mathrm{e,b}$ and the \ser\ index $n$ for 
both bulges of spiral galaxies and ellipticals. The range of 
$\mu_\mathrm{e,b}$ for different type objects is rather small 
(not greater than 5$^m$ for bulges and elliptical galaxies excluding 
dEs).

\item The Photometric Planes 1, 2, and 3 we considered, are 
relations between \ser\ model parameters which do not reveal 
anything new about bulges and elliptical galaxies. The PhP1 is a 
worse representation of the $\mu_\mathrm{0,b}$~--~$n$ self-correlation 
that arises as a consequence of not only using the \ser\ law, but also 
of the independence of the effective surface brightness from the 
\ser\ index and a small range of the effective surface brightness 
regardless of the objects considered. The PhP2 and PhP3 are another 
(noisy) version of the Kormendy relation and do not give any new 
information about the structure of bulges and elliptical galaxies. 

\item The Kormendy relation is a true correlation between main 
photometric parameters of bulges and elliptical galaxies which does 
help to divide galaxies into several populations. Classical bulges, 
pseudo-bulges, and elliptical galaxies have different origins and can 
not be described by curved relations proposed in 
\citet{grah2003,grah2011,grah2013a}. 
The bimodality of bulge parameters (or even trimodality, including 
elliptical galaxies) is a real observational fact.

We end by warning the readers when dealing with  correlations between quantities depending on hidden common parameters.

\end{enumerate}

\section*{Acknowledgments}
The authors express gratitude for the grant of the Russian 
Foundation for Basic Researches number 11-02-00471.
We are grateful to the referee for the comments and suggestions that 
lead to significant improvements of this paper. We thank Alister 
Graham and Dimitri Gadotti for useful and critical discussions.

This research has made use of the NASA/IPAC Extragalactic Database 
(NED) which is operated by the Jet Propulsion Laboratory, California 
Institute of Technology, under contract with the National 
Aeronautics and Space Administration. We made use of the LEDA 
database (http://leda.univ-lyon1.fr). We also acknowledge VizieR 
database (http://http://vizier.u-strasbg.fr/viz-bin/VizieR) as 
the main source of the samples used.

\bibliographystyle{mn2e}
\bibliography{art3}

\label{lastpage}

\end{document}